\documentclass[12pt,a4paper]{article}
\usepackage[dvips]{graphicx,color}
\usepackage{times}
\usepackage{xcolor}
\usepackage{cite}
\usepackage{float}
\usepackage[%
  colorlinks=true,
  urlcolor=blue,
  linkcolor=green,
  citecolor=blue]{hyperref}
\usepackage[utf8]{inputenc}
\usepackage{amsmath}
\usepackage{amsfonts}
\usepackage{makecell}
\usepackage{amssymb}
\usepackage{graphicx}
\usepackage{pgfplots}
\usepackage{subcaption}
\usepackage{caption}
\pgfplotsset{compat=1.18} 
\usepackage{multirow}
\usepackage{booktabs}
\usepackage{tabularray}
\UseTblrLibrary{booktabs}
\usepackage[left=2.0cm,right=2.0cm,top=2.5cm,bottom=2.5cm]{geometry}
\begin{document}
\newcommand{\sheptitle}
{Effects of RG running in breaking $\mu-\tau$ reflection symmetry, conventional versus minimal seesaw}
\newcommand{\shepauthor}
{Chandan Kumar Borah \footnote{ E-mail: cborah528@gmail.com}\ and\ 
 Chandan Duarah \footnote{ E-mail: chandanduarah@dibru.ac.in}}
\newcommand{\shepaddress}
   {Department of Physics, Dibrugarh University,
               Dibrugarh - 786004, India }
\newcommand{\shepabstract}
{The $\mu$--$\tau$ reflection symmetry has attracted considerable attention owing to its prediction of a maximal atmospheric mixing angle and a Dirac CP-violating phase $\delta=3\pi/2$, consistent with indications from the T2K and NO$\nu$A experiments. Since current neutrino oscillation data exhibit small but significant deviations from the exact symmetry predictions, understanding the origin of its breaking has become an important problem. In this work, we investigate the radiative breaking of $\mu$--$\tau$ reflection symmetry within the framework of the minimal seesaw model through the renormalization group (RG) evolution of neutrino parameters. Assuming the symmetry to be exact at the flavor symmetry (seesaw) scale, we examine whether the observed low-energy neutrino oscillation data can be reproduced after RG evolution. Owing to the rank-two structure of the minimal seesaw model, one light neutrino mass eigenvalue vanishes, reducing the number of independent high-energy neutrino parameters from five to four. We perform a systematic numerical analysis to determine the allowed high-energy parameter space capable of reproducing the current experimental constraints at low energies. Furthermore, a comparative study between the conventional Type-I seesaw and the minimal seesaw frameworks is carried out by analyzing the RG evolution of the solar and atmospheric mass-squared differences.

	\bigskip
	
	\noindent
	{\bfseries Keywords: Lepton mixing, $\mu-\tau$ reflection symmetry,  Renormalization group running, MSSM, MSM.} 
	\vspace{1.2cm}
\noindent
\pagebreak}
\begin{titlepage}
\begin{flushright}
\end{flushright}
\begin{center}
{\large{\bf\sheptitle}}
\bigskip\\
\shepauthor
\\
\mbox{}\\
{\it\shepaddress}\\
\vspace{.5in}
{\bf Abstract}
\bigskip
\end{center}
\setcounter{page}{0}
\shepabstract
\end{titlepage}

\section{Introduction}
\indent The experimentally observed equality in the magnitudes of the $\mu$- and $\tau$-flavor elements of the PMNS matrix \cite{PDG} suggests the possible existence of an underlying flavor symmetry \cite{RVW1, RVW2} governing neutrino mixing. Among the various symmetry patterns proposed in the literature, the $\mu$–$\tau$ permutation symmetry \cite{PS1, PS2} has attracted considerable attention for a long time. However, the experimental indication of a nearly maximal Dirac CP-violating phase, together with the equality of the moduli of the $\mu$- and $\tau$-flavor elements of the PMNS matrix, points toward another symmetry pattern. This symmetry, known as $\mu$–$\tau$ reflection symmetry \cite{RVW1}, can be realized in both the PMNS mixing matrix and the neutrino mass matrix. The symmetry transformation is given by

$$\nu_e \rightarrow \nu_e^c,\  \nu_\mu \rightarrow \nu_\tau^c,\  \nu_\tau \rightarrow \nu_\mu^c$$
and the corresponding neutrino mass matrix invariant under this transformation takes the form
\begin{equation}
    M=\begin{bmatrix}
        M_{ee} & M_{e\mu} & M_{e\tau}^* \\ M_{e\mu} & M_{\mu\mu} &M_{\mu\tau} \\ M_{e\tau}^* &M_{\mu\tau} & M_{\mu\mu}^*
    \end{bmatrix}.
\end{equation}
$\mu$–$\tau$ reflection symmetry predicts the CP-violating phase, $\delta = \pi/2$ or $3\pi/2$ and, this prediction is consistent with recent results from the T2K and NO$\nu$A experiments \cite{T2k1, T2k2, Nova1, Nova2}. The near maximal value of $\delta$ is also reflected in the recent global analysis of neutrino oscillation data \cite{GlobAnal} in the inverted order(IO) scenario, thereby bringing reflection symmetry into particular focus in current studies. Accordingly, a lot of studies have been conducted on this symmetry pattern in recent years \cite{RRS12, RRS4, RRS5, RRS6, RRS13, NNath, dirac, RRS9, VVV1, VVV2, RRS10, M.Kashav}.

The general features of the $\mu$--$\tau$ reflection symmetry have been extensively investigated in Refs.~\cite{RVW1,RVW2}. One of its most important implications is that it constrains the leptonic CP-violating phases and the atmospheric mixing angle $\theta_{23}$ to their maximal values. The specific predictions for the CP-violating phases, however, depend on the parametrization adopted for the lepton mixing matrix. In the standard parametrization, the Pontecorvo--Maki--Nakagawa--Sakata (PMNS) matrix is conventionally written as \cite{PDG}
\begin{equation}
U=P_1VP_2,
\label{U}
\end{equation}
where
\begin{equation}
    V=\begin{bmatrix}
  c_{12}c_{13}&s_{12}s_{13}&s_{13}e^{-i\delta}\\-s_{12}c_{23}-c_{12}s_{23}e^{i\delta}&c_{12}c_{23}-s_{12}s_{23}e^{i\delta}&c_{13}s_{23}\\s_{12}s_{23}-c_{12}s_{13}e^{i\delta}&-c_{12}s_{23}-s_{12}c_{23}e^{i\delta}&c_{13}c_{23} \end{bmatrix},
  \label{para}
\end{equation}
with $P_1=\mathrm{Diag}(e^{i\phi_1},e^{i\phi_2},e^{i\phi_3})$ and $P_2=\mathrm{Diag}(e^{i\rho},e^{i\sigma},1)$. Here, $s_{ij}=\sin\theta_{ij}$ and $c_{ij}=\cos\theta_{ij}$ ($ij=12,23,13$), where $\theta_{12}$, $\theta_{23}$ and $\theta_{13}$ denote the solar, atmospheric and reactor mixing angles, respectively, while $\delta$ is the Dirac CP-violating phase. The diagonal matrix $P_1$ contains the three unphysical phases $\phi_1$, $\phi_2$ and $\phi_3$, whereas $P_2$ contains the two Majorana CP phases $\rho$ and $\sigma$.

Ref.~\cite{RVW2} demonstrated that the generalized $\mu-\tau$ reflection symmetry can be realized through a suitable construction of the lepton mixing matrix U, without modifying its standard parametrization
\begin{equation}
U=A_{\mu\tau}U^{*}\zeta,
\label{Uconstraint}
\end{equation}
where, $A_{\mu\tau}$ is the $\mu$--$\tau$ exchange operator
\begin{equation}
A_{\mu\tau} = 
\begin{bmatrix}
1 & 0 & 0 \\
0 & 0 & 1 \\
0 & 1 & 0
\end{bmatrix},
\label{A}
\end{equation}
 and $\zeta=\mathrm{Diag} ({\eta_1,\eta_2,\eta_3})$ with $\eta_i=\pm1$. This defination in Eq. (\ref{Uconstraint}) of the mixing matrix directly leads to the general results of $\mu$--$\tau$ reflection symmetry that the atmospheric mixing angle $\theta_{23}=\pi/4$ and the Dirac CP phase $\delta=\pi/2$ or $3\pi/2$, while the unphysical phases and Majorana phases are restricted to take only discrete values, namely integer multiples of $\pi/2$. In this work, we adopt the Harrison--Scott (HS) formulation of $\mu-\tau$ reflection symmetry, in which the first-row elements of the lepton mixing matrix are chosen to be real, leading to a more restrictive phase structure \cite{RS1}. The corresponding mixing matrix is given by
\begin{equation}
    V_{HS}=\begin{bmatrix}
        u_1 & u_2 &u_3 \\ v_1 &v_2 &v_3 \\ v_1^* & v_2^* & v_3^*
    \end{bmatrix},
    \label{HS}
\end{equation}
where $u_i$ are real and $v_i$ are complex parameters. This parametrization automatically satisfies the condition $|U_{\mu i}|=|U_{\tau i}|$, and the mixing matrix remains invariant under the simultaneous interchange of the second and third rows followed by complex conjugation. To realise the HS formulation of $\mu-\tau$ reflection symmetry in the standard PMNS matrix given in Eq. \ref{U}, we have to consider some particular discrete values of the CP phases. Following Ref. \cite{CL1}, we consider two sets of values of the CP phases corresponding to two values of $\delta$:

\begin{itemize}
\item \textbf{Case I:} $\theta_{23} = \pi/4$, $\delta = \pi/2$, with
\begin{equation}
\alpha = \beta = \tfrac{3\pi}{2}, \quad \phi_1 = \tfrac{\pi}{2}, \quad \phi_2 = \phi_3 = 0,
\label{ch1}
\end{equation}

\item \textbf{Case II:} $\theta_{23} = \pi/4$, $\delta = 3\pi/2$, with
\begin{equation}
\alpha = \beta = \tfrac{\pi}{2}, \quad \phi_1 = \tfrac{3\pi}{2}, \quad \phi_2 = \phi_3 = 0.
\label{ch2}
\end{equation}
\end{itemize}
It is important to note that the above phase assignments reproduce the Harrison--Scott (HS) formulation in the standard parametrization of the lepton mixing matrix given in Eq.~(\ref{U}). These symmetry-induced constraints on the CP-violating phases constitute a key prediction of the $\mu$--$\tau$ reflection symmetry and will serve as the basis for the two distinct numerical analyses carried out in this work.

Although $\mu$–$\tau$ reflection symmetry predicts maximal values for certain neutrino mixing parameters, global analyses of oscillation data (Table \ref{GA}) demonstrate noticeable deviations from these predictions. These deviations motivate a detailed investigation of symmetry-breaking effects through various theoretical frameworks. Renormalization group (RG) running provides an effective framework for understanding the deviations from the exact predictions of $\mu$–$\tau$ reflection symmetry. The RG equations explicitly describe how the neutrino parameters evolve with the energy scale. Within this framework, the symmetry is assumed to hold exactly at a high-energy scale, such as the flavor symmetry or seesaw scale. As the parameters are evolved down to the electroweak scale through the RG equations, quantum corrections gradually break the symmetry, resulting in the observed deviations from maximality at low energies. The RG effects on neutrino parameters have been widely explored in the literature \cite{KS, RGE2, 1, R1, R3, R2, poko, babu, R4, R5, ohl, Dzhang, NNSingh, SGupta, RGERS2, YLZhou, JZhu, JMei, CBorah1, CBorah2, CBorah3, Pegu1, Pegu2}. In particular, several studies have investigated the influence of RG running in frameworks incorporating $\mu$–$\tau$ reflection symmetry \cite{CBorah1, CBorah2, CBorah3, RRS1, RRS6, dirac, JZhu, YLZhou, RGERS2}. The explicit energy-scale evolution of the neutrino parameters can be obtained by numerically solving the coupled RGEs. Starting from suitable initial conditions at the high-energy scale, the RG evolution yields predictions for the neutrino parameters at low energies. When $\mu$--$\tau$ reflection symmetry is imposed at the high-energy scale, the parameter space is characterized by five free parameters, namely the three neutrino mass eigenvalues together with the mixing angles $\theta_{12}$ and $\theta_{13}$. Determining a set of high-energy input parameters that reproduces all current experimental constraints after RG evolution is a computationally demanding problem. Such an optimization has, however, been accomplished in Refs.~\cite{CBorah1,CBorah2}.

In the present work, we incorporate the prediction of the minimal seesaw model (MSM) in radiative breaking of $\mu-\tau$ reflection symmetry. The key feature of MSM is that one mass eigenvalue vanishes under the theoretical setup. Although the concept of the Minimal Seesaw Model was proposed earlier \cite{Minimal}, but in recent years it has attracted considerable attention. In particular, this model has been extensively studied in the context of the radiative breaking of $\mu$–$\tau$ reflection symmetry \cite{RRS1, Minimal2, Minimal3}. The MSM describes physics above the mass scale of the lightest right-handed (RH) neutrino. In this energy regime, two RH neutrinos remain active. As the energy is lowered, these heavy states are sequentially integrated out, and at the scale of the lightest RH neutrino, one finally obtains the effective Weinberg operator ($\kappa$). Below the seesaw scale, the dimension-five Weinberg operator gives rise to the neutrino mass term. The RGEs of the neutrino parameters can be derived with the help of the RGE of this operator. After spontaneous symmetry breaking, this operator generates light Majorana neutrino masses, with the neutrino mass matrix being proportional to both $\kappa$ and the vacuum expectation value (vev) of the Higgs field. In this work, we treat the Higgs vev as an energy-dependent quantity.

\begin{table}[t]
\begin{center}
\begin{tabular}{c cc cc }
\hline
 Without SK atmospheric data & & \\ \hline
\multirow{2}{*}{Parameter}& 
\multicolumn{2}{c}{Normal Ordering}&
\multicolumn{2}{c}{Inverted Ordering} \\
\cline{2-5}
 &Best-fit Value  & $3\sigma$ & best-fit value & $3\sigma$ \\ \hline
 $\theta_{12}$         & 33.68    &  31.63-35.95   & 33.68    &  31.63-35.95         \\
 $\theta_{23}$         & 48.5    & 41.0-50.5       & 48.6    & 41.4-50.6      \\
 $\theta_{13}$         & 8.52    & 8.18-8.87        & 8.58    & 8.24-8.91    \\ 
 $\delta$         & 177     & 96-422  & 285    & 201-348         \\
 $\Delta m^2_{21}(/10^{-5}eV^2)$  & 7.49   & 6.92-8.05 & 7.49   & 6.92-8.05        \\
 $\Delta m^2_{32}(/10^{-3}eV^2)$  & 2.534 & 2.463-2.606 & -2.510     & -2.584-2.438          \\ \hline
  With SK atmospheric data & & \\ \hline
 $\theta_{12}$         & 33.68    &  31.63-35.95   & 33.68    &  31.63-35.95         \\
 $\theta_{23}$         & 43.3    & 41.3-49.9       & 47.9    & 41.5-49.8      \\
 $\theta_{13}$         & 8.56    & 8.19-8.89        & 8.59    & 8.25-8.93    \\ 
 $\delta$         & 212     & 124-364  & 274    & 201-335         \\
 $\Delta m^2_{21}(/10^{-5}eV^2)$  & 7.49   & 6.92-8.05 & 7.49   & 6.92-8.05        \\
 $\Delta m^2_{32}(/10^{-3}eV^2)$  & 2.513 & 2.451-2.578 & -2.484     & -2.547-2.421          \\ \hline
\end{tabular}
\end{center}
\caption{The best-fit values and 3$\sigma$ allowed ranges of neutrino oscillation parameters in NO and IO obtained from global analysis \cite{GlobAnal}.}
\label{GA}
\end{table}

Fixing one of the neutrino mass eigenvalues to zero at the seesaw scale, as predicted in the minimal seesaw model (MSM), reduces the number of free parameters at the high-energy scale to four. We therefore investigate whether this reduced parameter space, through renormalization group (RG) evolution, can successfully reproduce the observed low-energy neutrino oscillation data. It is worth noting that the vanishing neutrino mass eigenvalue remains unchanged throughout the RG evolution, reflecting a characteristic feature of the MSM. In addition to examining the RG evolution within the minimal seesaw framework, we perform a detailed comparison with the corresponding results obtained in the conventional Type-I seesaw scenario. The comparison is carried out by examining the low-energy predictions of the solar and atmospheric mass-squared differences, $\Delta m_{21}^{2}$ and $\Delta m_{32}^{2}$, obtained in the two frameworks after RG evolution from the common flavor symmetry scale. To quantify the differences between the two realizations, we introduce two relative ratios, $R_{21}$ and $R_{32}$, which measure the deviations of the solar and atmospheric mass-squared differences predicted in the minimal seesaw framework from those obtained in the conventional Type-I seesaw scenario. These ratios provide a simple quantitative measure of the extent to which the reduced parameter space of the MSM reproduces the low-energy phenomenology of the conventional Type-I seesaw framework. To the best of our knowledge, although extensive studies of the minimal seesaw model have been carried out in the literature, particularly in the context of neutrino masses, mixing, and leptogenesis, a systematic comparison of the RG evolution of neutrino parameters between the minimal seesaw and conventional Type-I seesaw frameworks has received little attention. Such a study is therefore well motivated, especially in view of the growing interest in the minimal seesaw model as a simple and economical framework capable of simultaneously explaining neutrino masses and providing a viable setting for leptogenesis.

We adopt the RGEs for the neutrino mixing parameters derived in our previous work~\cite{CBorah1} with a modification that one mass eigenvalue is taken to be zero now. In addition, we also consider the RGEs for the gauge and Yukawa couplings. These constitute a set of coupled differential equations, which we solve numerically using {\tt Python} code. Starting from the symmetry--predicted maximal values at the flavor symmetry scale $\Lambda_{FS}$, we evolve the parameters down to the electroweak scale $\Lambda_{EW}$, typically taken to be the top quark mass scale $m_t$. Below the seesaw scale, we work within the minimal supersymmetric standard model (MSSM). We assume supersymmetry is broken at some intermediate scale. 
Since the exact value of the SUSY breaking scale is not known, we perform our analysis for three different benchmark choices, namely $1~\text{TeV}$, $7~\text{TeV}$, and $14~\text{TeV}$. Below this scale, the effective theory is the Standard Model, where we additionally take into account the RGE for the Higgs quartic coupling.  

The remaining part of this paper is organized as follows. Section 2 discusses the minimal seesaw model and renormalization group evolution. In Section 3, we present the numerical analysis and results. Finally, the conclusions are given in Section 4.

\section{Conventional vs minimal seesaw and corresponding RG equations}

The Type-I seesaw mechanism is one of the most attractive and widely studied extensions of the Standard Model, providing a natural explanation for the tiny masses of the observed light neutrinos. In its conventional realization, three heavy right-handed neutrinos are introduced, one corresponding to each generation of fermions. After electroweak symmetry breaking, the exchange of these heavy Majorana neutrinos generates an effective Majorana mass matrix for the light neutrinos through the seesaw relation,
\begin{equation}
M_{\nu}=-M_D M_R^{-1}M_D^T,
\end{equation}
where $M_D$ and $M_R$ denote the $3\times3$ Dirac and Majorana neutrino mass matrices, respectively. The smallness of the light neutrino masses naturally follows from the heaviness of the right-handed neutrinos.

Neutrino oscillation experiments are sensitive to two independent neutrino mass-squared differences, defined as
\begin{equation}
\Delta m_{21}^{2}=m_{2}^{2}-m_{1}^{2},
\end{equation}
and either
\begin{equation}
|\Delta m_{32}^{2}|=|m_{3}^{2}-m_{2}^{2}|,
\end{equation}
or
\begin{equation}
|\Delta m_{31}^{2}|=|m_{3}^{2}-m_{1}^{2}|,
\end{equation}
where $m_1$, $m_2$, and $m_3$ are the three light neutrino mass eigenvalues. Since oscillation experiments determine only the absolute value of the atmospheric mass-squared difference, two possible neutrino mass orderings remain compatible with the current data, namely the normal ordering (NO),
\begin{equation}
m_1<m_2<m_3,
\end{equation}
and the inverted ordering (IO),
\begin{equation}
m_3<m_1<m_2.
\end{equation}
Although oscillation experiments do not determine the absolute neutrino mass scale, cosmological observations place a stringent upper bound on the sum of neutrino masses,
\begin{equation}
\sum_i m_i < 0.12~\mathrm{eV},
\end{equation}
which provides an important constraint on viable neutrino mass models~\cite{mbound}.

As the present oscillation data establish only two independent neutrino mass-squared differences, the existence of three heavy right-handed neutrinos is not a phenomenological necessity. This observation motivates the consideration of the minimal seesaw model (MSM), in which only two heavy right-handed neutrinos are introduced. Despite its reduced particle content, the MSM successfully accounts for the observed neutrino oscillation phenomena while retaining the essential features of the conventional Type-I seesaw mechanism.

In the minimal seesaw framework, the Dirac and Majorana neutrino mass matrices are of dimensions $3\times2$ and $2\times2$, respectively. Consequently, the effective light-neutrino mass matrix has rank two, implying that one of the three light neutrino mass eigenvalues vanishes identically. Accordingly,
\begin{equation}
m_1=0 \qquad \text{for NO},
\end{equation}
whereas
\begin{equation}
m_3=0 \qquad \text{for IO}.
\end{equation}

In the present work, these conditions are imposed at the flavor symmetry scale,
\begin{equation}
\Lambda_{\rm FS}=10^{14}~\mathrm{GeV},
\end{equation}
where the $\mu$--$\tau$ reflection symmetry is assumed to hold exactly. Since our analysis is restricted to energies below the seesaw scale, the heavy right-handed neutrinos are integrated out of the theory and the low-energy effective dynamics is described by the dimension-five Weinberg operator. Consequently, the renormalization group evolution is governed by the same effective RGEs as in the conventional Type-I seesaw framework below the seesaw scale, and the vanishing neutrino mass eigenvalue remains unchanged throughout the RG evolution.

An immediate consequence of the rank-two structure of the minimal seesaw model is the reduction in the number of independent neutrino parameters at the flavor symmetry scale. Under exact $\mu$--$\tau$ reflection symmetry, the conventional Type-I seesaw framework contains five free neutrino parameters, whereas the minimal seesaw framework contains only four because one neutrino mass eigenvalue is fixed to zero. A central objective of the present work is therefore to investigate whether this reduced parameter space is sufficient to reproduce the observed low-energy neutrino oscillation data after renormalization group evolution.

Since the heavy right-handed neutrinos are integrated out below the seesaw scale, the renormalization group evolution of the neutrino masses, leptonic mixing angles, and CP-violating phases is governed by the same one-loop RGEs as those in the conventional Type-I seesaw framework. The complete set of one-loop RGEs for the effective neutrino mass operator, gauge couplings, Yukawa couplings, Higgs quartic coupling, and all neutrino parameters in the MSSM and SM effective theories has been derived in our previous work~\cite{CBorah1}. Therefore, these equations are not reproduced here for brevity.

The only modification introduced in the present analysis concerns the high-energy boundary conditions. At $\Lambda_{\rm FS}$, one neutrino mass eigenvalue is fixed to zero according to the minimal seesaw hypothesis, while the remaining neutrino masses, leptonic mixing angles, and CP-violating phases satisfy the $\mu$--$\tau$ reflection symmetry conditions. Starting from these boundary conditions, the coupled RGEs are solved numerically from the flavor symmetry scale down to the electroweak scale to obtain the low-energy neutrino parameters.

Finally, the resulting low-energy predictions are compared with the latest global-fit neutrino oscillation data. Furthermore, to assess the impact of reducing the number of right-handed neutrinos, we perform a direct comparison between the minimal seesaw and the conventional Type-I seesaw frameworks by examining their predictions for the solar and atmospheric mass-squared differences. For this purpose, two dimensionless quantities, $R_{21}$ and $R_{32}$, are introduced in the following section to quantify the relative differences between the low-energy predictions of the two scenarios.

\section{Numerical analysis and results}

In this section, we investigate the renormalization group (RG) running behaviour of neutrino parameters and its role in inducing the breaking of exact $\mu$–$\tau$ reflection symmetry within the framework of the minimal seesaw framework. We assume that the exact symmetry is realized at a high-energy flavor symmetry scale, $\Lambda_{\rm FS}=10^{14}  \mathrm{GeV}$, motivated by the typical mass scale of heavy right-handed neutrinos. Employing the RGEs for neutrino parameters derived in our previous work \cite{CBorah1}, with the additional constraint that one light neutrino mass eigenvalue vanishes, we evolve the parameters from $\Lambda_{\rm FS}$ down to the electroweak scale and examine the deviations from the exact symmetry predictions induced by RG effects. At $\Lambda_{\rm FS}$, the CP-violating phases are assigned their maximal values as predicted by $\mu$–$\tau$ reflection symmetry. The gauge and Yukawa coupling constants are obtained from their low-energy values through a bottom-up RG evolution, following the procedure described in Ref.~\cite{CBorah1}. In the minimal seesaw framework, one neutrino mass eigenvalue is fixed to zero, leaving two nonzero mass eigenvalues together with the mixing angles $\theta_{12}$ and $\theta_{13}$ as free input parameters at $\Lambda_{\rm FS}$. During the RG evolution, we consider three representative values of the SUSY-breaking scale, $\Lambda_s$, and implement the appropriate matching conditions at this scale as discussed in Ref.~\cite{CBorah1}.

\begin{table}[t]
\centering
\begin{tabular}{c c c c c}
\hline
\textbf{Parameter} & \textbf{Input value at $m_t$ scale} & \multicolumn{3}{c}{\textbf{Output value at $\Lambda_{FS}$ for different $\Lambda_s$}} \\ \cline{3-5}
              &            & \textbf{$1TeV$} & \textbf{$7TeV$} & \textbf{$14TeV$} \\ \hline
$g_1$         & 0.46125    & 0.633482  &  0.625793  & 0.623121         \\ 
$g_2$         &0.66239     & 0.702064  & 0.684943   & 0.679140         \\ 
$g_3$         & 1.18955    & 0.745271   & 0.721910  & 0.715024         \\ 
$y_t$         & 0.9917     & 0.763613   & 0.700936 &  0.685736         \\ 
$y_b$         & 0.01571      & 0.679651   & 0.601777  & 0.583075         \\ 
$y_{\tau}$    & 0.01006    & 0.779141  & 0.735524   & 0.725049         \\ \hline
\end{tabular}
\caption{Input values of gauge and Yukawa coupling at $m_t=172GeV$ and corresponding output values at $\Lambda_{FS}=10^{14}GeV$ for three different values of $\Lambda_s$.} 
\label{gandy}
\end{table}

The present analysis is motivated by two main objectives. First, compared to our previous study based on the conventional Type-I seesaw framework \cite{CBorah1}, where five free neutrino parameters were specified at the high-energy scale, the minimal seesaw scenario contains only four free parameters due to the vanishing of one neutrino mass eigenvalue. It is therefore important to investigate whether a suitable set of these four high-scale parameters can successfully reproduce the current experimental constraints at low energies through RG evolution. Second, we aim to compare the RG effect on neutrino parameters in the minimal seesaw framework with those obtained in the conventional Type-I seesaw scenario. Such a comparison provides insight into the impact of the reduced parameter space on the radiative breaking of $\mu$–$\tau$ reflection symmetry and the resulting low-energy neutrino phenomenology.

The entire analysis is carried out separately for the two possible neutrino mass orderings, namely the normal ordering (NO) and the inverted ordering (IO). Furthermore, each mass-ordering scenario is divided into two distinct categories, denoted as Case I and Case II, corresponding to different choices of the CP-violating phases. This classification enables a systematic investigation of the RG-induced breaking of $\mu$–$\tau$ reflection symmetry under different CP-phase configurations. In each case, we consider $\tan\beta=58$.

To quantify the difference of the effects induced by the RG evolution between the Type-I seesaw and minimal seesaw frameworks, we define the following dimensionless ratios:
\begin{equation}
R_{21}=\frac{\Delta m_{21,\rm Type-I}^{2}-\Delta m_{21,\rm MS}^{2}}{\Delta m_{21,\rm Type-I}^{2}},
\end{equation}
and
\begin{equation}
R_{32}=\frac{\Delta m_{32,\rm Type-I}^{2}-\Delta m_{32,\rm MS}^{2}}{\Delta m_{32,\rm Type-I}^{2}},
\end{equation}
which represent the relative differences in the solar and atmospheric mass-squared differences, respectively, with the Type-I seesaw results taken as the reference. These quantities provide a quantitative measure of the relative deviation in the RG evolution between the two seesaw frameworks. The values of $R_{21}$ and $R_{32}$ are evaluated for the different neutrino mass orderings, CP-phase configurations, and 
\begin{table}[t]
\begin{center}
\begin{tabular}{c cc cc cc}
\hline
\multirow{2}{*}{Parameter}& 
\multicolumn{2}{c}{$\Lambda_s=1$TeV}&
\multicolumn{2}{c}{$\Lambda_s=7$TeV}&
\multicolumn{2}{c}{$\Lambda_s=14$TeV} \\
\cline{2-7}
 &\makecell{Input at \\ $\Lambda_{FS}$}  & \makecell{Output at \\$\Lambda_{EW}$} &\makecell{Input at \\ $\Lambda_{FS}$}&\makecell{Output at \\ $\Lambda_{EW}$ }&\makecell{Input at \\ $\Lambda_{FS}$}&\makecell{ Output at \\ $\Lambda_{EW}$} \\ \hline
 $m_1\ (eV)$& 0  & 0 & 0 & 0 & 0&  0 \\
 $m_2\ (eV)$& 0.006467 & 0.008746 & 0.006467 & 0.00872 & 0.006467  &  0.008706 \\
 $m_3\ (eV)$& 0.036338 &  0.053492 & 0.036338 & 0.05239 & 0.036338 & 0.051935 \\
$\theta_{13} (/^\circ)$&8.5685 & 8.56014 & 8.57 & 8.5632 & 8.57 & 8.5637\\
$\theta_{12} (/^\circ)$& 34.00044 & 33.9931 & 34 & 33.9931 & 34 & 33.9933 \\
$\theta_{23} (/^\circ)$& 45 & 45.0754 & 45 & 45.0629 & 45 & 45.0589 \\
 $\delta (/^\circ)$& 90 & 89.9478 & 90 & 89.9590 & 90 & 89.9625 \\
 $\Delta m^2_{21}(10^{-5}eV^2)$&- & 7.64 & - & 7.62 &- & 7.57 \\
 $\Delta m^2_{32}(10^{-3}eV^2)$&- & 2.78 & - & 2.66 &- & 2.62 \\
 $\sum_i m_i (eV)$&- & 0.06223 & - & 0.061120 &- &  0.06064 \\
 \hline
\end{tabular}
\end{center}
\caption{Input values at $\Lambda_{FS}$ and corresponding low energy values at $m_t$ scale of all the parameters for three different values of $\Lambda_s=1, 7$ and $14$ TeV in NO and case-I. }
\label{TNO1}
\end{table}
\begin{table}[!h]
\begin{center}
\begin{tabular}{c cc cc cc}
\hline
\multirow{2}{*}{Parameter}& 
\multicolumn{2}{c}{$\Lambda_s=1$TeV}&
\multicolumn{2}{c}{$\Lambda_s=7$TeV}&
\multicolumn{2}{c}{$\Lambda_s=14$TeV} \\
\cline{2-7}
 &\makecell{Input at \\ $\Lambda_{FS}$}  & \makecell{Output at \\$\Lambda_{EW}$} &\makecell{Input at \\ $\Lambda_{FS}$}&\makecell{Output at \\ $\Lambda_{EW}$ }&\makecell{Input at \\ $\Lambda_{FS}$}&\makecell{ Output at \\ $\Lambda_{EW}$} \\ \hline
 $m_1\ (eV)$& 0  & 0 & 0 & 0 & 0&  0 \\
 $m_2\ (eV)$& 0.006467 & 0.008640 & 0.006467 & 0.008559 & 0.006467  &  0.008517 \\
 $m_3\ (eV)$& 0.036338 &  0.050804 & 0.036338 & 0.05154 & 0.036338 & 0.05122 \\
$\theta_{13} (/^\circ)$&8.5598 & 8.5619 & 8.57 & 8.5619 & 8.57 & 8.5626\\
$\theta_{12} (/^\circ)$& 33.9508 & 33.9931 & 34 & 33.9614 & 34 & 33.9648 \\
$\theta_{23} (/^\circ)$& 45 & 45.0426 & 45 & 45.0349 & 45 & 45.0324 \\
 $\delta (/^\circ)$& 270 & 270.0158 & 270 &  270.0125 &270 & 270.0114 \\
 $\Delta m^2_{21}(10^{-5}eV^2)$&- & 7.46 & - & 7.32 &- & 7.25 \\
 $\Delta m^2_{32}(10^{-3}eV^2)$&- & 2.50 & - & 2.58 &- & 2.55 \\
 $\sum_i m_i (eV)$&- & 0.05944 & - & 0.060106 &- &  0.059728 \\
 \hline
\end{tabular}
\end{center}
\caption{Input values at $\Lambda_{FS}$ and corresponding low energy values at $m_t$ scale of all the parameters for three different values of $\Lambda_s=1, 7$ and $14$ TeV in NO and case-II. }
\label{TNO2}
\end{table}
representative SUSY-breaking scales considered in this work. Table~\ref{R} summarizes the values of the relative differences $R_{21}$ and $R_{32}$ between the Type-I seesaw and minimal seesaw frameworks at the flavor symmetry scale, $\Lambda_{\rm FS}$, and the electroweak scale, $\Lambda_{\rm EW}$, for the NO and IO scenarios under Cases I and II of the CP phases. The results are presented for the three representative SUSY-breaking scales, $\Lambda_s=1,\ 7,$ and $14$ TeV, thereby providing a quantitative measure of the differences between the two seesaw realizations. For a more transparent comparison, the corresponding results are also illustrated by means of bar charts in Figs.~\ref{FR1}--\ref{FR4}.

\subsection{The NO scenario}
In this case, the lightest mass eigenvalue is $m_1$ and as per the minimal seesaw case, we will consider $m_1=0$. Therefore, we have the high-energy free parameter set as ($m_2,m_3,\theta_{12},\theta_{13}$). 

\begin{table}[ht]
\centering
\label{tab:ratio}
\footnotesize
\setlength{\tabcolsep}{3pt}

\resizebox{\textwidth}{!}{%
\begin{tblr}{
    width=\textwidth,
    colspec={Q[c,wd=1.2cm] Q[c,wd=2cm] *{6}{Q[c]}},
    row{1,2}={font=\bfseries},
    hline{1,Z}={1pt},
    hline{2}={0.6pt},
    hline{3,5,7,9}={0.5pt},
    cell{3}{1}={r=2}{},
    cell{5}{1}={r=2}{},
    cell{7}{1}={r=2}{},
    cell{9}{1}={r=2}{},
}

&
Parameter
& \SetCell[c=2]{c} $\Lambda_s=1~\mathrm{TeV}$
& &
\SetCell[c=2]{c} $\Lambda_s=7~\mathrm{TeV}$
& &
\SetCell[c=2]{c} $\Lambda_s=14~\mathrm{TeV}$
& \\

&
&
At $\Lambda_{FS}$
&  $\Lambda_{EW}$
&  $\Lambda_{FS}$
&  $\Lambda_{EW}$
&  $\Lambda_{FS}$
&  $\Lambda_{EW}$ \\

\rotatebox{90}{NO Case I}
& $R_{21}$ &-5.141& 24.813& 1.695&25.525 & 3.606 & 25.525 \\
& $R_{32}$ &-9.513 &-11.629 & -6.149& -7.749& -4.228& -5.636\\

\rotatebox{90}{NO Case II}
& $R_{21}$ & 2.047&2.752 & 0.663&2.752 &0.183 &2.752 \\
& $R_{32}$ &0.496 & 1.660& -2.866&-0.406 &-1.228 &1.616 \\

\rotatebox{90}{IO Case I}
& $R_{21}$ &6.838 & 28.512&5.590 & -0.785&0.042 &2.229 \\
& $R_{32}$ &-1.839 &1.333 & -4.516& -3.930&-2.121 &-1.418 \\

\rotatebox{90}{IO Case II}
& $R_{21}$ &-7.288 &-16.127 & 0.552&25.059 &0.266 &26.549 \\
& $R_{32}$ &-1.243 &-2.728 &-4.132 & -0.727& -5.150& -1.943\\

\end{tblr}}
\caption{Relative differences between the Type-I seesaw and minimal seesaw predictions for the solar and atmospheric mass-squared differences, quantified through the ratios $R_{21}$ and $R_{32}$ at the flavor symmetry scale $\Lambda_{\rm FS}$ and the electroweak scale $\Lambda_{\rm EW}$ for the NO and IO scenario under Cases I and II of the CP phases. The comparison is presented for three representative SUSY-breaking scales, $\Lambda_s=1,\ 7,$ and $14$ TeV.}
\label{R}
\end{table}

The chosen input values of the free parameters and the corresponding low-energy output values are presented in Table~\ref{TNO1}. The CP phases are fixed according to Eq.~\ref{ch1}; however, to avoid redundancy and save space, their explicit input and output values are not listed in the table. For the Case~I scenario, the resulting low-energy neutrino parameters, shown in Table~\ref{TNO1}, are all found to lie within the $3\sigma$ ranges of the current global-fit data presented in Table \ref{GA}. Although the deviations of $\theta_{23}$ from $45^\circ$ and $\delta$ from its symmetry-predicted value are relatively small, they provide clear evidence of the radiative breaking of the $\mu$--$\tau$ reflection symmetry. Therefore, the renormalization group evolution successfully generates symmetry-breaking effects while maintaining consistency with the experimentally allowed $3\sigma$ ranges of all neutrino oscillation parameters.

Similarlly, for NO of mass eigenvalues and case II for the CP phases  we have presented the results in Table \ref{TNO2}. In this scenario also we are successful in obtaining the high energy values for the four free parameter which can reproduce the experimental constraints within the $3\sigma$ range of global analysis data.

After achieving the primary objective of identifying a set of four free parameters capable of successfully reproducing the experimental constraints at low energies, we perform a comparative analysis between the conventional seesaw and minimal seesaw scenarios. For consistency, the input values of the free parameters and the remaining model parameters are chosen to be the same as those used in Ref.~\cite{CBorah1} for the conventional seesaw case. To avoid unnecessary repetition, these numerical values are not tabulated here. Instead, our focus is on highlighting the similarities and differences between the predictions of the two frameworks.

\begin{figure}[!t]
\begin{tabular}{cc}
   \begin{subfigure}[b]{0.8\textwidth}
    \centering
    \includegraphics[height=5cm]{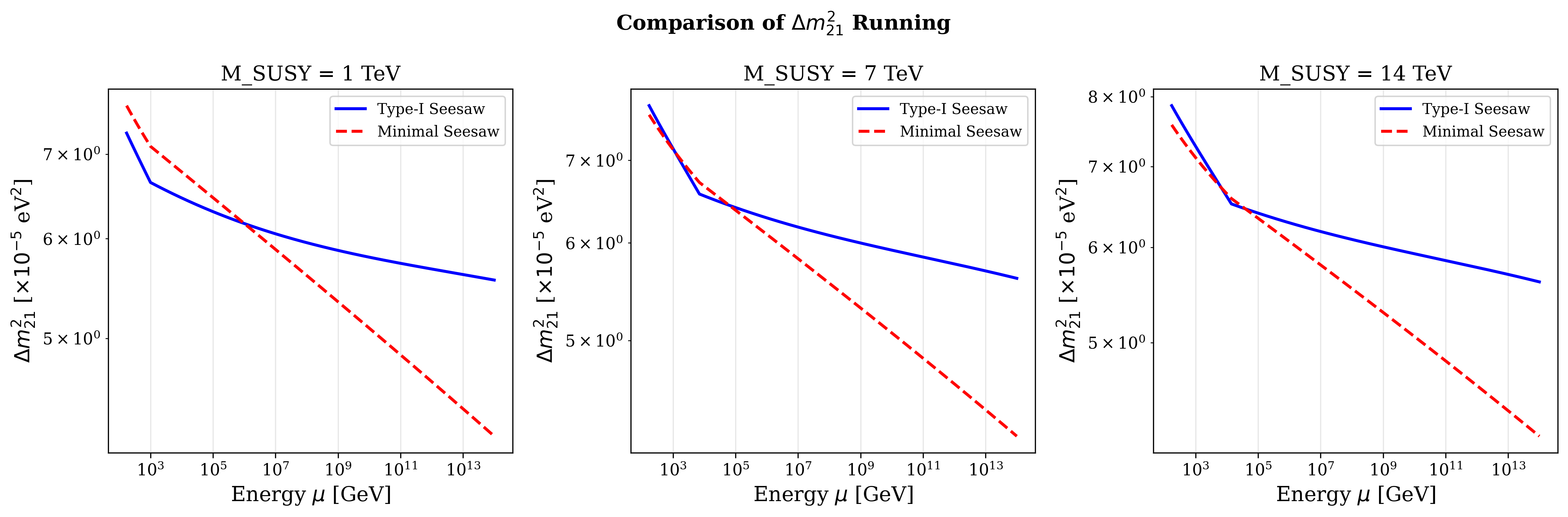}
    \label{m1_NO1}
  \end{subfigure}\\[2ex]
\begin{subfigure}[b]{0.8\textwidth}
    \centering
    \includegraphics[height=5cm]{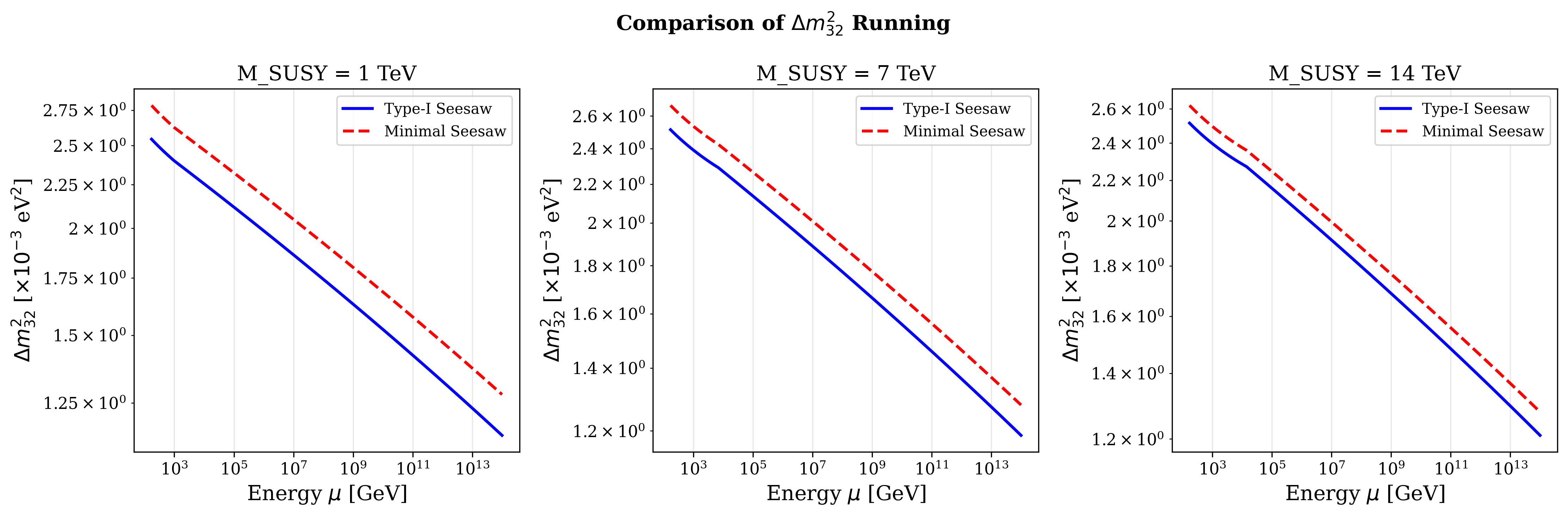}
    \label{m2_NO1}
  \end{subfigure} 
    \end{tabular}
  \caption{Comparison of the RG running of the mass squared differences in minimal and type I seesaw for NO and case-I with three different values of $\Lambda_s$. The blue and red lines represent the RG running in type I seesaw and minimal seesaw frameworks respectively.}
  \label{FNO1}
\end{figure}

\begin{figure}[!t]
\begin{tabular}{cc}
   \begin{subfigure}[b]{\textwidth}
    \centering
    \includegraphics[height=5cm]{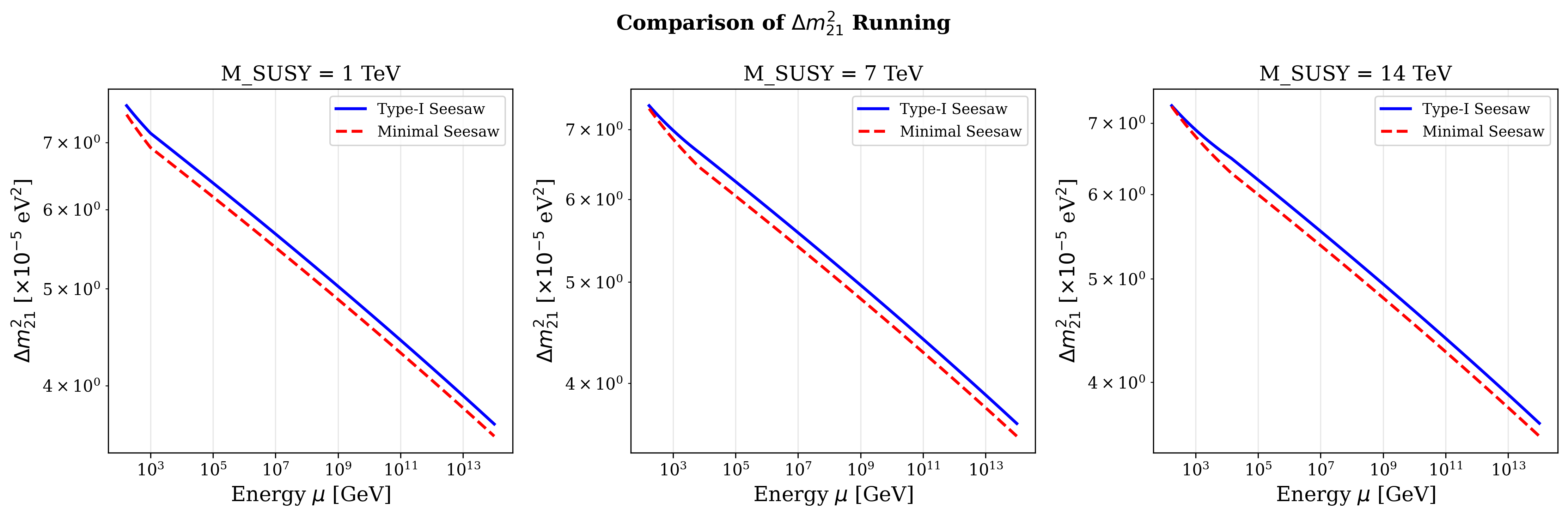}
    \label{m1_NO2}
  \end{subfigure}\\[2ex]
\begin{subfigure}[b]{\textwidth}
    \centering
    \includegraphics[height=5cm]{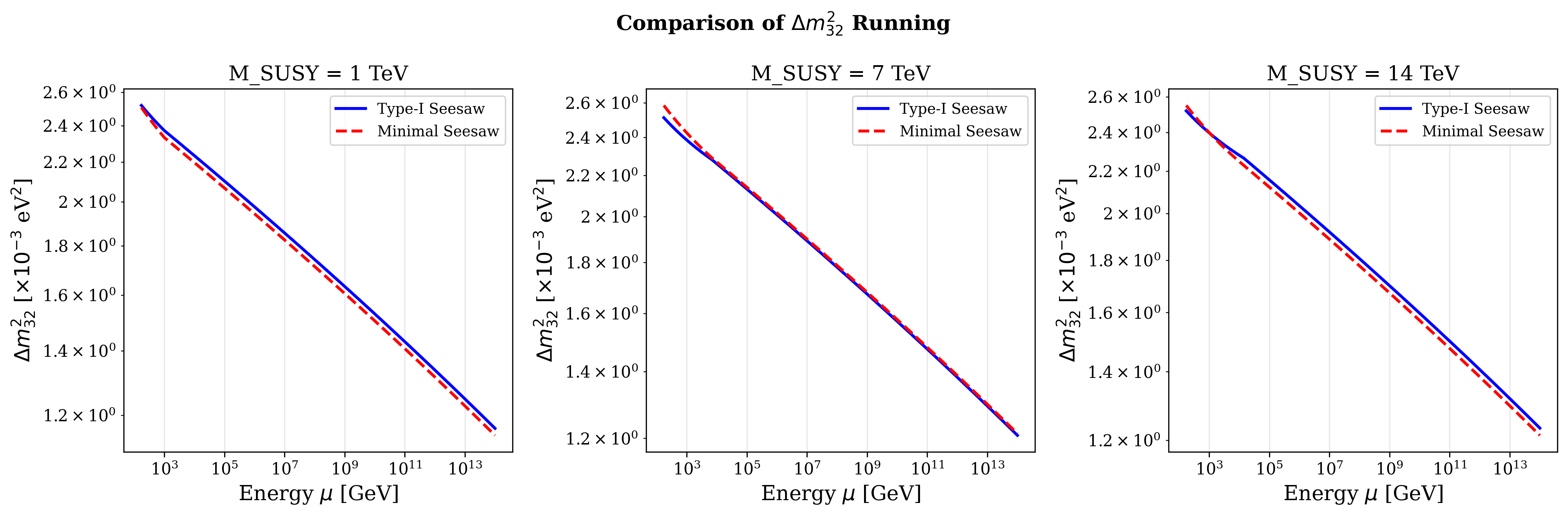}
    \label{m2_NO2}
  \end{subfigure} 
    \end{tabular}
  \caption{Comparison of the RG running of the mass squared differences in minimal and type I seesaw for NO and case-II with three different values of $\Lambda_s$. The blue and red lines represent the RG running in type I seesaw and minimal seesaw frameworks respectively.}
  \label{FNO2}
\end{figure}

Furthermore, since the renormalization group evolution of all neutrino parameters has already been discussed in detail in Ref.~\cite{CBorah1}, we restrict our presentation to the energy evolution of the mass-squared differences, which provides a clear and direct comparison between the conventional seesaw and minimal seesaw formulations. The comparative running behaviors of the mass-squared differences are displayed in Figs.~\ref{FNO1} and \ref{FNO2}, enabling a direct assessment of the impact of reducing the number of right-handed neutrinos from three to two on the radiative evolution of neutrino masses.

\subsubsection{Case I of NO} 
\textbf{Analyses of $\Delta m_{21}^2$:}
The RG running behaviour of the neutrino mass-squared differences $\Delta m_{21}^{2}$ and $\Delta m_{32}^{2}$ for the NO scenario with Case I of the CP phases in the Type-I seesaw and minimal seesaw frameworks is presented in Fig.~\ref{FNO1}. The first three panels compare the RG evolution of $\Delta m_{21}^{2}$, while the remaining three panels illustrate the corresponding evolution of $\Delta m_{32}^{2}$ for three representative SUSY-breaking scales, $\Lambda_s=1,\ 7,$ and $14$ TeV.

As shown in the first three panels, $\Delta m_{21}^{2}$ increases monotonically as the energy scale evolves from the flavor symmetry scale $\Lambda_{\rm FS}$ down to the electroweak scale $\Lambda_{\rm EW}$ in both the Type-I seesaw and minimal seesaw scenarios. This behaviour reflects the cumulative effect of radiative corrections on the effective neutrino mass matrix during the RG evolution. Although both models exhibit qualitatively similar running behaviour, noticeable quantitative differences emerge as the energy approaches the flavor symmetry scale. In particular, the numerical evolution of $\Delta m_{21}^{2}$ in the minimal seesaw framework differs noticeably from that obtained in the conventional Type-I seesaw model, leading to an increasing separation between the two curves as the energy scale approaches the flavor symmetry scale. Another noteworthy feature is that this qualitative behaviour remains essentially unchanged for all three choices of the SUSY-breaking scale considered in the present analysis. Although varying $\Lambda_s$ produces a slight shift in the numerical values of $\Delta m_{21}^{2}$, the relative evolution of the two models remains nearly identical. This indicates that the observed difference between the Type-I seesaw and minimal seesaw scenarios is largely insensitive to the choice of the SUSY-breaking scale within the parameter space considered. It should be noted that the comparison presented here is performed using the respective phenomenologically viable parameter spaces of the two models. Consequently, the quantitative differences observed in Fig.~\ref{FNO1} arise from the combined effects of their distinct theoretical structures and the corresponding sets of input parameters. Nevertheless, the results clearly demonstrate that the RG evolution of the solar mass-squared difference exhibits distinguishable characteristics in the Type-I seesaw and minimal seesaw frameworks.

\textbf{Analyses of $\Delta m_{32}^2$:} The last three panels of Fig.~\ref{FNO1} compare the RG evolution of the atmospheric mass-squared difference, $\Delta m_{32}^{2}$, in the Type-I seesaw and minimal seesaw frameworks for $\Lambda_s=1,\ 7,$ and $14$ TeV. Similar to the behaviour observed for $\Delta m_{21}^{2}$, the value of $\Delta m_{32}^{2}$ increases monotonically as the energy scale evolves from the flavor symmetry scale $\Lambda_{\rm FS}$ to the electroweak scale $\Lambda_{\rm EW}$ in both frameworks. This behaviour reflects the continuous modification of the effective neutrino mass matrix due to radiative corrections during the RG evolution. A comparison of the two models shows that the RG running of $\Delta m_{32}^{2}$ is qualitatively very similar over the entire energy range considered. However, the minimal seesaw scenario consistently predicts slightly larger values of $\Delta m_{32}^{2}$ than the conventional Type-I seesaw model, resulting in an approximately uniform separation between the two curves throughout the RG evolution. In contrast to the behaviour of $\Delta m_{21}^{2}$, no significant enhancement of the separation is observed as the energy approaches the flavor symmetry scale. This indicates that the radiative evolution of the atmospheric mass-squared difference is comparatively less sensitive to the differences between the two seesaw realizations. It is also observed that varying the SUSY-breaking scale from $\Lambda_s=1$ TeV to $14$ TeV has only a marginal impact on the running behaviour in either framework. The overall evolution and the relative difference between the Type-I seesaw and minimal seesaw scenarios remain essentially unchanged for all three choices of $\Lambda_s$. As in the case of $\Delta m_{21}^{2}$, the numerical comparison is performed within the phenomenologically allowed parameter spaces of the respective models. Therefore, the observed differences should be regarded as the combined consequence of their distinct theoretical structures and corresponding input parameters rather than being attributed solely to the underlying seesaw mechanism. 
\begin{figure}[!t]
    \centering
    \includegraphics[width=1 \textwidth]{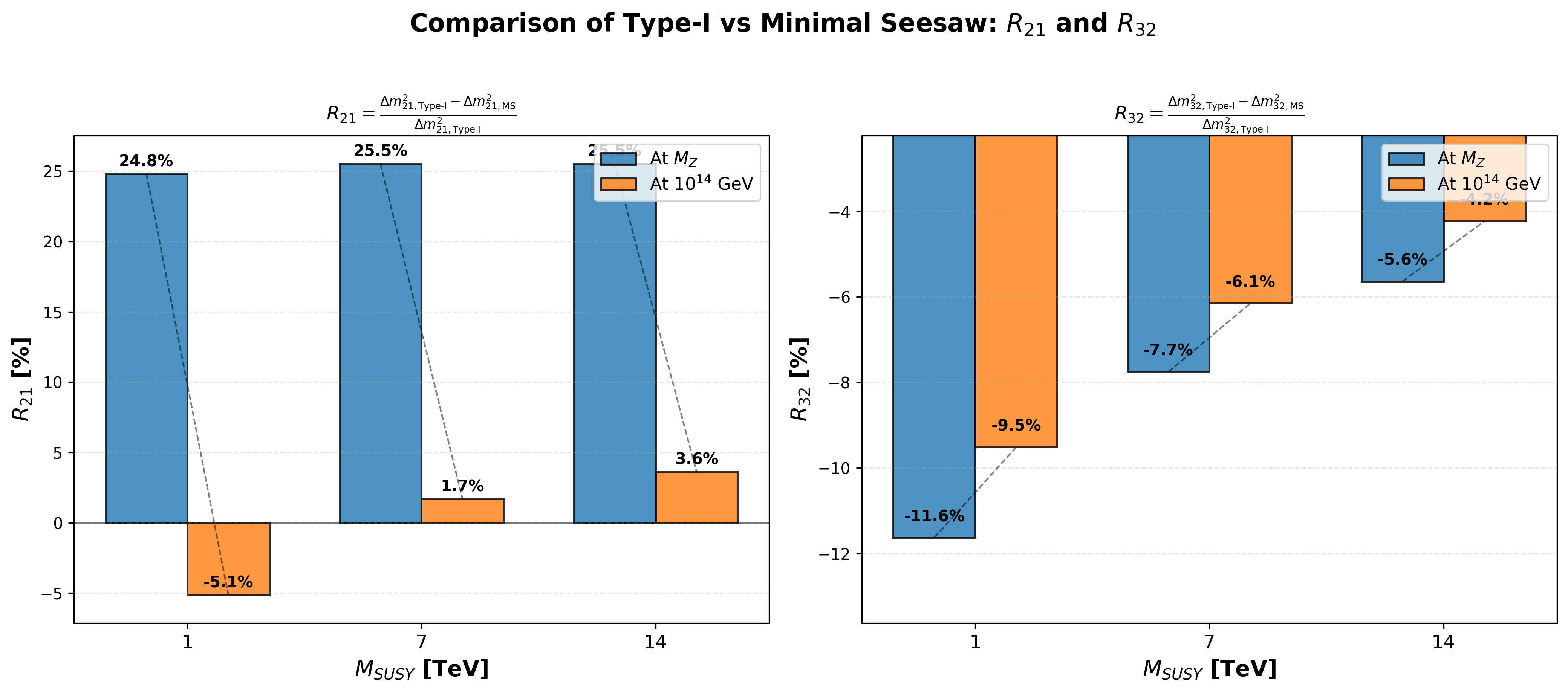}
    \caption{Comparison of $R_{21}$ (left panel) and $R_{32}$ (right panel) between the Type-I seesaw and minimal seesaw frameworks for the NO and Case I. The blue and orange bars correspond to the values at $\Lambda_{\rm EW}$, and the $\Lambda_{\rm FS}=10^{14}$ GeV, respectively, for three SUSY-breaking scales, $\Lambda_s=1,\ 7,$ and $14$ TeV.}
    \label{FR1}
\end{figure}

Overall, the comparison indicates that the RG evolution of the solar mass-squared difference, $\Delta m_{21}^{2}$, exhibits a relatively stronger model dependence than that of the atmospheric mass-squared difference, $\Delta m_{32}^{2}$, suggesting that $\Delta m_{21}^{2}$ is more sensitive to the underlying seesaw realization within the considered parameter space.

\textbf{Analyses of the ratio R:} For the NO scenario with Case I of the CP phases, the values of $R_{21}$ exhibit a strong dependence on the energy scale. At the flavor symmetry scale, $R_{21}$ changes from a negative value at $\Lambda_s=1$ TeV to positive values at $\Lambda_s=7$ and $14$ TeV, indicating that the relative difference between the Type-I seesaw and minimal seesaw predictions is significantly reduced as the SUSY-breaking scale increases. In contrast, at the electroweak scale, $R_{21}$ remains consistently large and positive for all three values of $\Lambda_s$, demonstrating that the accumulated radiative corrections lead to sizeable differences in the low-energy predictions of the two frameworks. The corresponding values of $R_{32}$ remain negative at both $\Lambda_{\rm FS}$ and $\Lambda_{\rm EW}$ for all three SUSY-breaking scales. Furthermore, the magnitude of $R_{32}$ gradually decreases with increasing $\Lambda_s$, suggesting that the atmospheric mass-squared difference becomes progressively less sensitive to the underlying seesaw realization as the SUSY-breaking scale increases.

Figure~\ref{FR1} presents the relative differences $R_{21}$ and $R_{32}$ between the Type-I seesaw and minimal seesaw frameworks for the NO spectrum with Case I of the CP phases. The blue and orange bars represent the values of the corresponding ratios evaluated at the electroweak scale, $\Lambda_{\rm EW}$, and the flavor symmetry scale, $\Lambda_{\rm FS}=10^{14}$ GeV, respectively, for the three representative SUSY-breaking scales, $\Lambda_s=1,\ 7,$ and $14$ TeV.

The left panel illustrates the behaviour of $R_{21}$. At the electroweak scale, the relative difference remains remarkably stable, with values close to $25\%$ for all three choices of $\Lambda_s$. This indicates that the low-energy predictions of the solar mass-squared difference in the two seesaw frameworks differ by nearly one quarter, irrespective of the SUSY-breaking scale. In contrast, the corresponding values at the flavor symmetry scale exhibit a pronounced dependence on $\Lambda_s$. For $\Lambda_s=1$ TeV, $R_{21}$ is negative, indicating that the minimal seesaw prediction exceeds that of the Type-I seesaw framework. As the SUSY-breaking scale increases to $7$ TeV and $14$ TeV, $R_{21}$ becomes positive, while its magnitude decreases to only a few percent. This behaviour demonstrates that the predictions of the two seesaw frameworks gradually converge at higher SUSY-breaking scales. The right panel shows the corresponding comparison for $R_{32}$. In contrast to $R_{21}$, the values of $R_{32}$ remain negative at both $\Lambda_{\rm EW}$ and $\Lambda_{\rm FS}$ for all three values of $\Lambda_s$, indicating that the atmospheric mass-squared difference predicted by the minimal seesaw framework remains consistently larger than that obtained in the Type-I seesaw scenario. Furthermore, the magnitude of $R_{32}$ decreases steadily with increasing $\Lambda_s$ at both energy scales, revealing that the discrepancy between the two frameworks becomes progressively smaller as the SUSY-breaking scale increases. Overall, Fig.~\ref{FR1} quantitatively confirms the conclusions drawn from the RG evolution plots. The relative difference associated with the solar mass-squared difference, $R_{21}$, is considerably larger than that of the atmospheric mass-squared difference, $R_{32}$, indicating that $\Delta m_{21}^{2}$ is substantially more sensitive to the underlying seesaw realization. Moreover, increasing the SUSY-breaking scale generally reduces the relative difference between the Type-I seesaw and minimal seesaw predictions, particularly at the flavor symmetry scale.

\begin{figure}[t]
    \centering
    \includegraphics[width= \textwidth]{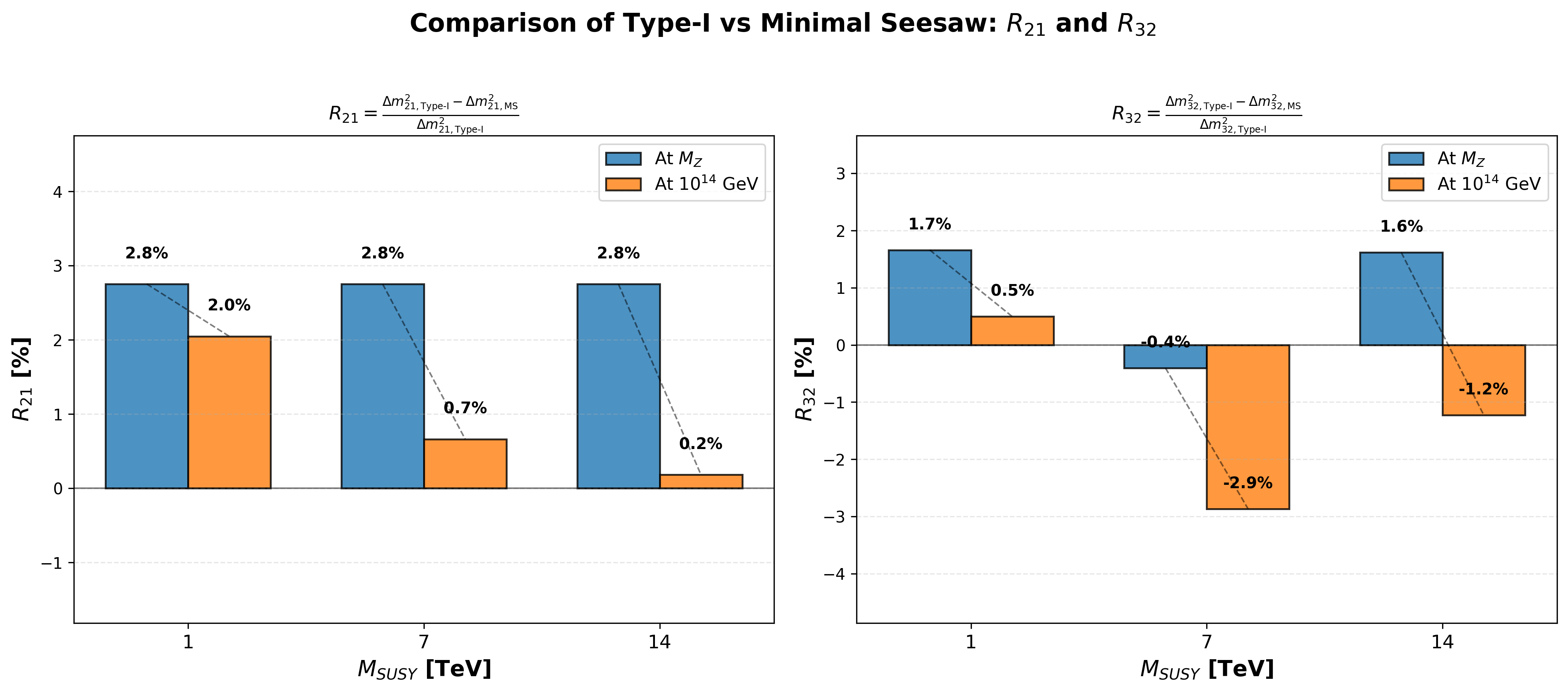}
    \caption{Comparison of $R_{21}$ (left panel) and $R_{32}$ (right panel) between the Type-I seesaw and minimal seesaw frameworks for the NO and Case II. The blue and orange bars correspond to the values at $\Lambda_{\rm EW}$, and the $\Lambda_{\rm FS}=10^{14}$ GeV, respectively, for three SUSY-breaking scales, $\Lambda_s=1,\ 7,$ and $14$ TeV.}
    \label{FR2}
\end{figure}

\subsubsection{Case II of NO}
We have also carried out the corresponding comparative analysis for the NO spectrum with Case II of the CP phases. The resulting comparison plots are shown in Fig.~\ref{FNO2}. Similar to Fig.~\ref{FNO1}, the first three panels compare the RG evolution of $\Delta m_{21}^{2}$, while the remaining three panels present the corresponding comparison for $\Delta m_{32}^{2}$ at $\Lambda_s=1,\ 7,$ and $14$ TeV. The overall RG running behaviour and the relative evolution of the Type-I seesaw and minimal seesaw scenarios remain qualitatively similar to those observed in Fig.~\ref{FNO1}. Although the numerical values differ owing to the different choice of CP phases, the comparison once again demonstrates that the two frameworks exhibit distinct radiative evolution while preserving the same qualitative features over the entire energy range considered.

\textbf{Analyses of $\Delta m_{21}^2$:} The comparison of the RG evolution of the solar mass-squared difference, $\Delta m_{21}^{2}$, for the NO spectrum with Case II of the CP phases is presented in the first three panels of Fig.~\ref{FNO2}. Similar to the Case I scenario, $\Delta m_{21}^{2}$ increases monotonically as the energy scale evolves from the flavor symmetry scale $\Lambda_{\rm FS}$ to the electroweak scale $\Lambda_{\rm EW}$ in both the Type-I seesaw and minimal seesaw frameworks. This behaviour reflects the continuous radiative modification of the effective neutrino mass matrix during the RG evolution. Unlike the behaviour observed in Fig.~\ref{FNO1}, the Type-I seesaw framework predicts slightly larger values of $\Delta m_{21}^{2}$ than the minimal seesaw scenario throughout the entire energy range. The two curves exhibit nearly identical slopes and remain approximately parallel, indicating that the relative difference between the two frameworks is almost independent of the energy scale. Consequently, no appreciable enhancement in the separation between the two curves is observed as the energy approaches the flavor symmetry scale. It is further observed that the qualitative behaviour remains unchanged for all three representative SUSY-breaking scales, $\Lambda_s=1,\ 7,$ and $14$ TeV. Although the numerical values of $\Delta m_{21}^{2}$ undergo small changes with the variation of $\Lambda_s$, the overall RG evolution and the relative difference between the Type-I seesaw and minimal seesaw frameworks remain essentially unaffected. This demonstrates that the comparison is robust against variations of the SUSY-breaking scale within the considered parameter space.

As in the Case I analysis, the numerical comparison is carried out using the respective phenomenologically viable parameter spaces of the two models. Therefore, the quantitative differences observed in Fig.~\ref{FNO2} should be interpreted as arising from the combined effects of the distinct theoretical structures of the two seesaw frameworks and their corresponding input parameter sets. Nevertheless, the results indicate that the two frameworks exhibit distinguishable RG evolution of the solar mass-squared difference while preserving the same qualitative running behaviour.

\textbf{Analyses of $\Delta m_{32}^2$:} The last three panels of Fig.~\ref{FNO2} present the comparison of the RG evolution of the atmospheric mass-squared difference, $\Delta m_{32}^{2}$, in the Type-I seesaw and minimal seesaw frameworks for the three representative SUSY-breaking scales, $\Lambda_s=1,\ 7,$ and $14$ TeV. Similar to $\Delta m_{21}^{2}$, the value of $\Delta m_{32}^{2}$ increases monotonically as the energy scale evolves from the flavor symmetry scale $\Lambda_{\rm FS}$ to the electroweak scale $\Lambda_{\rm EW}$ in both frameworks, reflecting the cumulative effect of radiative corrections on the effective neutrino mass matrix. It is observed that the RG evolution of $\Delta m_{32}^{2}$ is remarkably similar in the two seesaw scenarios. The corresponding curves exhibit nearly identical slopes over the entire energy range, with only a very small numerical difference between the Type-I seesaw and minimal seesaw predictions. In particular, for $\Lambda_s=7$ TeV, the two curves almost overlap, whereas for $\Lambda_s=1$ TeV and $14$ TeV only a slight separation is visible. This indicates that the atmospheric mass-squared difference is comparatively less sensitive to the differences between the two seesaw realizations than the solar mass-squared difference. Furthermore, the variation of the SUSY-breaking scale has only a marginal influence on the comparison. Although small numerical shifts are observed as $\Lambda_s$ varies from $1$ TeV to $14$ TeV, the overall RG evolution and the relative behaviour of the Type-I seesaw and minimal seesaw frameworks remain essentially unchanged. As in the previous analyses, the comparison is performed using the respective phenomenologically viable parameter spaces of the two models. Therefore, the small numerical differences observed in Fig.~\ref{FNO2} arise from the combined effects of the distinct model structures and their corresponding input parameters, while the qualitative running behaviour remains practically identical in the two frameworks.

A comparison between Case I and Case II reveals that the choice of CP-phase configuration has a noticeable impact on the RG evolution of the neutrino mass-squared differences in both the Type-I seesaw and minimal seesaw frameworks. Although the qualitative running behaviour remains unchanged in all cases, the quantitative evolution differs depending on the adopted CP-phase assignment. For the solar mass-squared difference, $\Delta m_{21}^{2}$, the distinction between the two seesaw frameworks is more pronounced in Case I than in Case II. In Case I, the separation between the Type-I seesaw and minimal seesaw predictions gradually increases towards the flavor symmetry scale, indicating a relatively stronger dependence of the RG evolution on the underlying seesaw realization. In contrast, for Case II, the two curves remain nearly parallel throughout the entire energy range, with an almost constant separation. This suggests that the choice of CP phases in Case II suppresses the relative difference in the radiative evolution of $\Delta m_{21}^{2}$ between the two frameworks. A similar trend is observed for the atmospheric mass-squared difference, $\Delta m_{32}^{2}$. While both cases exhibit nearly identical RG running in the Type-I seesaw and minimal seesaw scenarios, the agreement is even stronger in Case II. In particular, for $\Lambda_s=7$ TeV, the two curves almost completely overlap, indicating that the RG evolution of $\Delta m_{32}^{2}$ becomes nearly insensitive to the underlying seesaw framework. By comparison, a slightly larger but still modest separation is observed in Case I.

The above comparison indicates that different CP-phase assignments associated with the $\mu$--$\tau$ reflection symmetric boundary conditions lead to different patterns of radiative corrections. Consequently, the extent of RG-induced breaking of the $\mu$--$\tau$ reflection symmetry depends not only on the underlying seesaw framework but also on the initial CP-phase configuration adopted at the flavor symmetry scale.

\textbf{Analyses of the ratio R:} For the NO scenario with Case II of the CP phases, both $R_{21}$ and $R_{32}$ are considerably smaller in magnitude than those obtained in Case I. The values of $R_{21}$ remain positive over the entire parameter space considered, while their magnitudes decrease systematically with increasing $\Lambda_s$, indicating a closer agreement between the Type-I seesaw and minimal seesaw predictions at higher SUSY-breaking scales. The behaviour of $R_{32}$ is comparatively stable, with only small positive or negative values observed. This demonstrates that the atmospheric mass-squared difference is only weakly affected by the choice of the seesaw framework in this CP-phase configuration.

Figure~\ref{FR2} presents the relative differences $R_{21}$ and $R_{32}$ between the Type-I seesaw and minimal seesaw frameworks for the NO spectrum with Case II of the CP phases. The blue and orange bars correspond to the values evaluated at the electroweak scale, $\Lambda_{\rm EW}$, and the flavor symmetry scale, $\Lambda_{\rm FS}=10^{14}$ GeV, respectively, for the three representative SUSY-breaking scales, $\Lambda_s=1,\ 7,$ and $14$ TeV.

The left panel shows the variation of $R_{21}$. At the electroweak scale, the relative difference remains nearly constant at approximately $2.8\%$ for all three choices of $\Lambda_s$, indicating that the low-energy predictions of the solar mass-squared difference obtained in the two seesaw frameworks are in excellent agreement. At the flavor symmetry scale, however, $R_{21}$ decreases systematically from about $2.0\%$ for $\Lambda_s=1$ TeV to only $0.2\%$ for $\Lambda_s=14$ TeV. This gradual reduction demonstrates that the high-energy predictions of the Type-I seesaw and minimal seesaw frameworks converge as the SUSY-breaking scale increases. The right panel presents the corresponding comparison for $R_{32}$. At the electroweak scale, the relative differences remain very small, varying between approximately $-0.4\%$ and $1.7\%$, thereby indicating an almost identical prediction for the atmospheric mass-squared difference in the two frameworks. At the flavor symmetry scale, the values of $R_{32}$ also remain close to zero, although both positive and negative values are observed depending on the choice of $\Lambda_s$. The sign changes simply indicate which of the two frameworks predicts a slightly larger value of $|\Delta m_{32}^{2}|$, while the small magnitudes confirm that the discrepancy between the two models is negligible.

\begin{table}[!t]
\begin{center}
\begin{tabular}{c cc cc cc}
\hline
\multirow{2}{*}{Parameter}& 
\multicolumn{2}{c}{$\Lambda_s=1$TeV}&
\multicolumn{2}{c}{$\Lambda_s=7$TeV}&
\multicolumn{2}{c}{$\Lambda_s=14$TeV} \\
\cline{2-7}
 &\makecell{Input at \\ $\Lambda_{FS}$}  & \makecell{Output at \\$\Lambda_{EW}$} &\makecell{Input at \\ $\Lambda_{FS}$}&\makecell{Output at \\ $\Lambda_{EW}$ }&\makecell{Input at \\ $\Lambda_{FS}$}&\makecell{ Output at \\ $\Lambda_{EW}$} \\ \hline
 $m_1\ (eV)$& 0  &0.04992 & 0 & 0.04973 & 0& 0.04957 \\
 $m_2\ (eV)$& 0.006467 & 0.05062 & 0.006467 & 0.050449 & 0.006467  &  0.0050275 \\
 $m_3\ (eV)$& 0 &  0 & 0 & 0 & 0 & 0 \\
$\theta_{13} (/^\circ)$&8.5685 & 8.5102 & 8.57 & 8.5781 & 8.57 & 8.5773\\
$\theta_{12} (/^\circ)$& 34.00044 & 33.3336 & 34 & 34.1698 & 34 & 34.4585 \\
$\theta_{23} (/^\circ)$& 45 & 44.9449 & 45 & 44.9516 & 45 & 44.9540 \\
 $\delta (/^\circ)$& 90 & 93.1769 & 90 & 93.1047 & 90 & 92.9774 \\
 $\Delta m^2_{21}(10^{-5}eV^2)$&- & 7.04 & - & 7.20 &- & 7.02 \\
 $\Delta m^2_{32}(10^{-3}eV^2)$&- & 2.56 & - & 2.54 &- & 2.52 \\
 $\sum_i m_i (eV)$&- & 0.100545 & - & 0.100178 &- &  0.099847 \\
 \hline
\end{tabular}
\end{center}
\caption{Input values at $\Lambda_{FS}$ and corresponding low energy values at $m_t$ scale of all the parameters for three different values of $\Lambda_s=1, 7$ and $14$ TeV in IO and case-I. }
\label{TIO1}
\end{table}

\begin{table}[!t]
\begin{center}
\begin{tabular}{c cc cc cc}
\hline
\multirow{2}{*}{Parameter}& 
\multicolumn{2}{c}{$\Lambda_s=1$TeV}&
\multicolumn{2}{c}{$\Lambda_s=7$TeV}&
\multicolumn{2}{c}{$\Lambda_s=14$TeV} \\
\cline{2-7}
 &\makecell{Input at \\ $\Lambda_{FS}$}  & \makecell{Output at \\$\Lambda_{EW}$} &\makecell{Input at \\ $\Lambda_{FS}$}&\makecell{Output at \\ $\Lambda_{EW}$ }&\makecell{Input at \\ $\Lambda_{FS}$}&\makecell{ Output at \\ $\Lambda_{EW}$} \\ \hline
 $m_1\ (eV)$& 0  & 0.048975 & 0 & 0.049848 & 0&  0.04988 \\
 $m_2\ (eV)$& 0.006467 & 0.04978 & 0.006467 & 0.050614 & 0.006467  &  0.050646 \\
 $m_3\ (eV)$& 0 &  0 & 0 & 0 & 0 & 0 \\
$\theta_{13} (/^\circ)$&8.5598 & 8.6906 & 8.57 & 8.8474 & 8.57 & 8.8467\\
$\theta_{12} (/^\circ)$& 34.9696 & 34.9696 & 34 & 36.071 & 34 & 36.1200 \\
$\theta_{23} (/^\circ)$& 45 & 44.9399 & 45 & 44.9584 & 45 & 44.9626 \\
 $\delta (/^\circ)$& 270 & 273.1307 & 270 &  270.0125 &270 & 269.7086 \\
 $\Delta m^2_{21}(10^{-5}eV^2)$&- & 8.02 & - & 7.68 &- & 7.67 \\
 $\Delta m^2_{32}(10^{-3}eV^2)$&- & 2.47 & - & 2.56 &- & 2.56 \\
 $\sum_i m_i (eV)$&- & 0.0987637 & - &  0.100462 &- &  0.100528 \\
 \hline
\end{tabular}
\end{center}
\caption{Input values at $\Lambda_{FS}$ and corresponding low energy values at $m_t$ scale of all the parameters for three different values of $\Lambda_s=1, 7$ and $14$ TeV in IO and case-II. }
\label{TIO2}
\end{table}
Overall, Fig.~\ref{FR2} demonstrates that the relative differences between the Type-I seesaw and minimal seesaw frameworks are substantially smaller than those obtained for the NO spectrum with Case I of the CP phases. In particular, both $R_{21}$ and $R_{32}$ remain within only a few percent over the entire parameter space considered. These results quantitatively support the conclusions drawn from the RG evolution plots, namely that the two seesaw frameworks exhibit nearly identical radiative evolution for the NO spectrum with Case II of the CP phases, especially at higher SUSY-breaking scales.

\subsection{Inverted Ordering}
Having completed the analysis for the normal ordering (NO) of neutrino masses in both Case I and Case II of the CP-phase configurations, we now turn to the corresponding analysis for the inverted ordering (IO) scenario. The numerical results obtained for Case I and Case II are presented in Tables~\ref{TIO1} and \ref{TIO2}, respectively.

As in the NO case, we restrict our discussion of the renormalization group evolution to the mass-squared differences. The explicit running behaviors of all neutrino parameters within the type-I seesaw framework have already been presented in our previous work, Ref.~\cite{CBorah1}. Since the minimal seesaw extension exhibits qualitatively similar running patterns, we do not reproduce those plots here. Instead, we focus on the evolution of the mass-squared differences, which allows for a concise and transparent comparison between the two frameworks.

\begin{figure}[t]
\begin{tabular}{cc}
   \begin{subfigure}[b]{\textwidth}
    \centering
    \includegraphics[height=5cm]{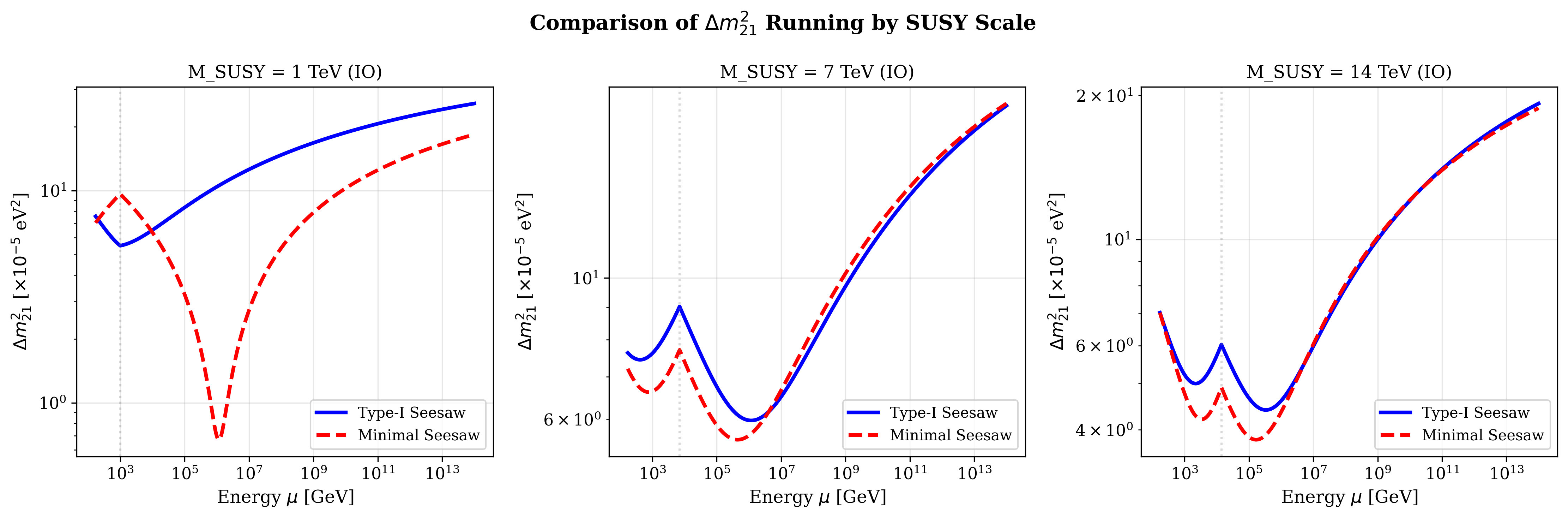}
    \label{m1_IO1}
  \end{subfigure}\\[2ex]
\begin{subfigure}[b]{\textwidth}
    \centering
    \includegraphics[height=5cm]{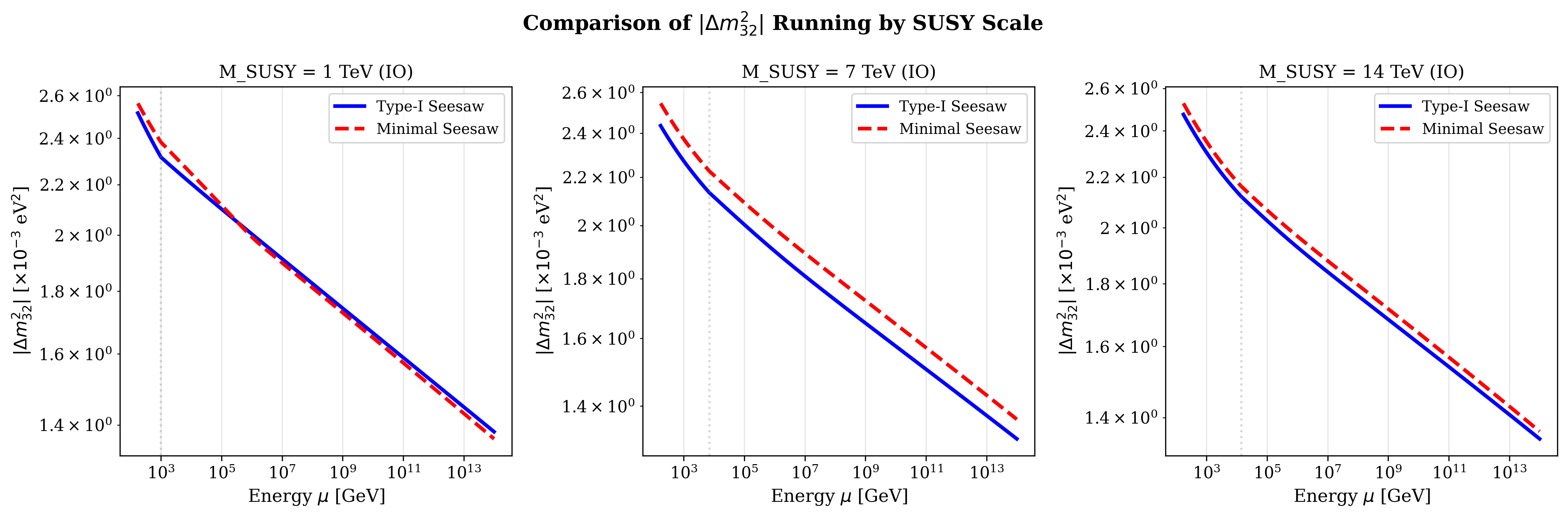}
    \label{m2_IO1}
  \end{subfigure} 
    \end{tabular}
  \caption{Comparison of the RG running of the mass squared differences in minimal and type I seesaw for IO and case-I with three different values of $\Lambda_s$.  The blue and red lines represent the RG running in type I seesaw and minimal seesaw frameworks respectively.}
  \label{FIO1}
\end{figure}

\begin{figure}[t]
    \centering
    \includegraphics[width= \textwidth]{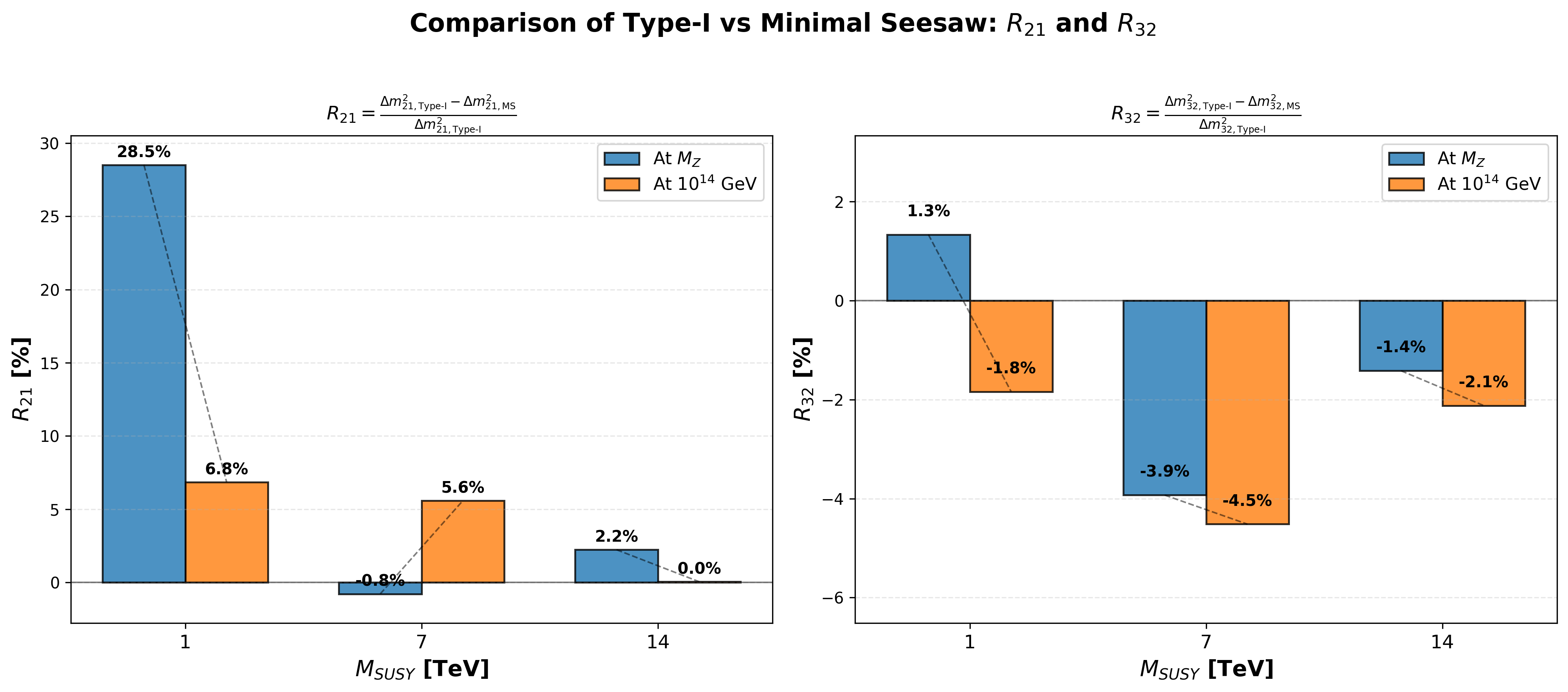}
    \caption{Comparison of $R_{21}$ (left panel) and $R_{32}$ (right panel) between the Type-I seesaw and minimal seesaw frameworks for the IO and Case I. The blue and orange bars correspond to the values at $\Lambda_{\rm EW}$, and the $\Lambda_{\rm FS}=10^{14}$ GeV, respectively, for three SUSY-breaking scales, $\Lambda_s=1,\ 7,$ and $14$ TeV.}
    \label{FR3}
\end{figure}

In the minimal seesaw scenario, the lightest neutrino mass eigenvalue is taken to be $m_3=0$, leaving four free parameters at the high-energy scale. The CP-phase values are chosen according to Eqs.~\ref{ch1} and \ref{ch2}, corresponding to the Case I and Case II configurations, respectively. For the chosen set of input values of the free parameters, the low-energy values of the neutrino mass-squared differences are found to lie within the $3\sigma$ ranges of the latest global-fit data.

\begin{figure}[t]
\begin{tabular}{cc}
   \begin{subfigure}[b]{0.8\textwidth}
    \centering
    \includegraphics[height=5cm]{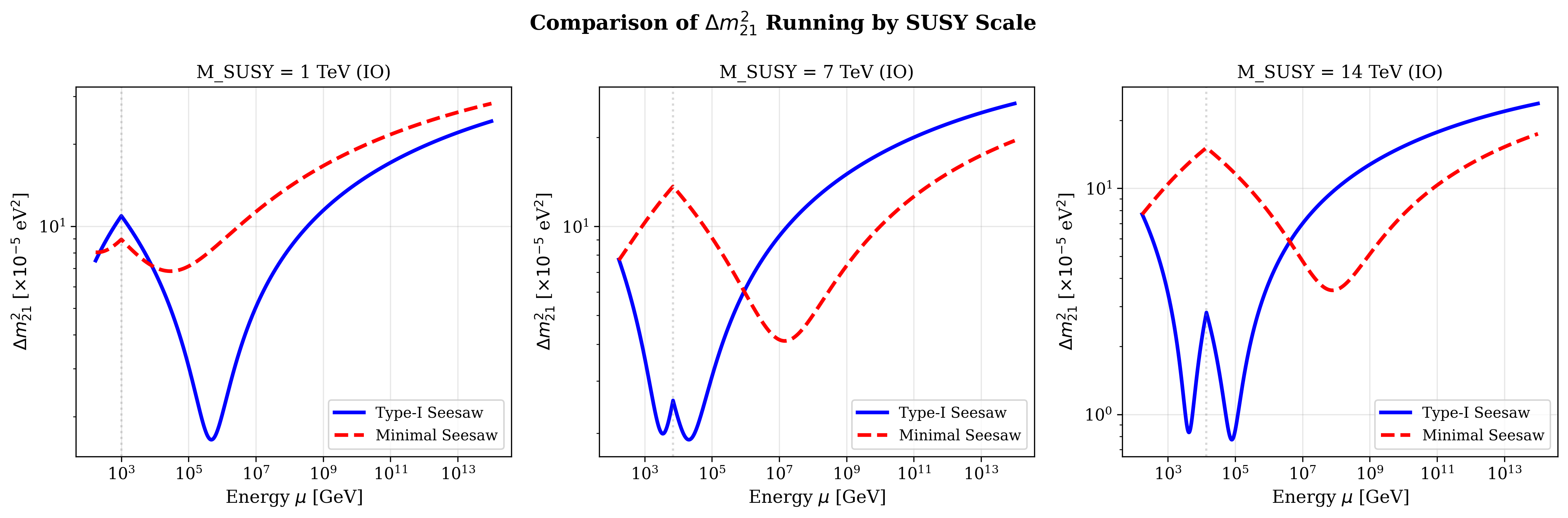}
    \label{m1_IO2}
  \end{subfigure}\\[2ex]
\begin{subfigure}[b]{0.8\textwidth}
    \centering
    \includegraphics[height=5cm]{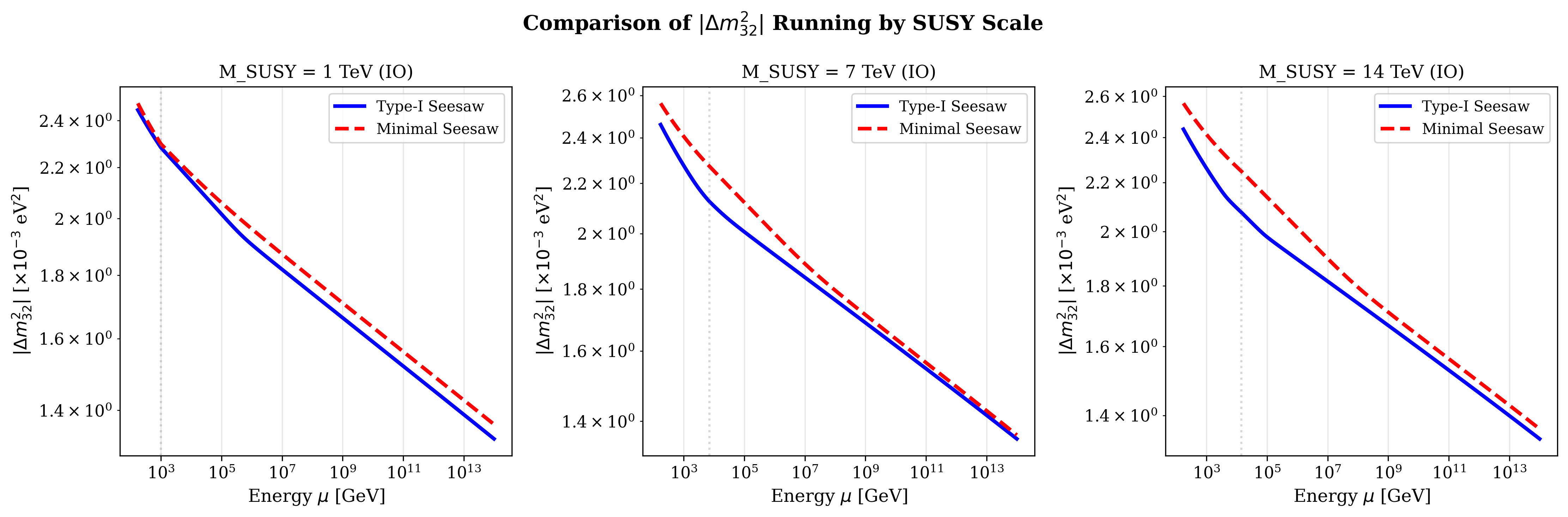}
    \label{m2_IO2}
  \end{subfigure} 
    \end{tabular}
  \caption{Comparison of the RG running of the mass squared differences in minimal and type I seesaw for IO and case-II with three different values of $\Lambda_s$. The blue and red lines represent the RG running in type I seesaw and minimal seesaw frameworks respectively.}
  \label{FIO2}
\end{figure}

The RG running behaviour of the mass-squared differences for the IO spectrum is presented in Figs.~\ref{FIO1} and \ref{FIO2}. Figure~\ref{FIO1} illustrates the comparison of the RG evolution of $\Delta m_{21}^{2}$ and $\Delta m_{23}^{2}$ in the Type-I seesaw and minimal seesaw frameworks for Case I of the CP phases. Similar to the NO analysis, the first three panels compare the running of $\Delta m_{21}^{2}$, while the remaining three panels present the corresponding comparison for $\Delta m_{23}^{2}$, each for the three representative SUSY-breaking scales, $\Lambda_s=1,\ 7$ and $14$ TeV.

\subsubsection{Case I of IO} 
\textbf{Analyses of $\Delta m_{21}^2$:}
The first three panels of Fig.~\ref{FIO1} compare the RG evolution of the solar mass-squared difference, $\Delta m_{21}^{2}$, in the Type-I seesaw and minimal seesaw frameworks for the IO spectrum with Case I of the CP phases. Unlike the NO scenario, the running behaviour of $\Delta m_{21}^{2}$ in the IO spectrum is non-monotonic at low energies. In both frameworks, $\Delta m_{21}^{2}$ initially decreases with increasing energy scale, reaches a minimum around the intermediate energy region, and subsequently increases continuously towards the flavor symmetry scale. This behaviour reflects the non-trivial interplay among the radiative corrections governing the evolution of the effective neutrino mass matrix. A noticeable difference between the two seesaw frameworks is observed for all three representative SUSY-breaking scales. For $\Lambda_s=1$ TeV, the minimal seesaw framework exhibits a significantly stronger suppression of $\Delta m_{21}^{2}$ around the intermediate energy region than the conventional Type-I seesaw model, resulting in a substantially larger separation between the two curves. As the SUSY-breaking scale is increased to $\Lambda_s=7$ TeV and $14$ TeV, the discrepancy between the two models gradually decreases, and the corresponding curves approach each other over most of the energy range. In particular, near the flavor symmetry scale, the predictions of the two frameworks become nearly identical for $\Lambda_s=7$ TeV and $14$ TeV. The figure also indicates that the SUSY-breaking scale has a more pronounced impact on the RG evolution of $\Delta m_{21}^{2}$ in the IO scenario than in the NO scenario. While the qualitative running behaviour remains unchanged, increasing $\Lambda_s$ noticeably reduces the numerical difference between the Type-I seesaw and minimal seesaw predictions. This suggests that, for the chosen parameter space, the radiative evolution of the solar mass-squared difference becomes progressively less sensitive to the underlying seesaw realization as the SUSY-breaking scale increases.

As in the previous analyses, the comparison is carried out using the respective phenomenologically viable parameter spaces of the two frameworks. Therefore, the observed quantitative differences should be interpreted as resulting from the combined effects of the distinct model structures and their corresponding input parameters rather than from the underlying seesaw mechanism alone.

\textbf{Analyses of $\Delta m_{32}^2$:} The last three panels of Fig.~\ref{FIO1} compare the RG evolution of the atmospheric mass-squared difference, $|\Delta m_{32}^{2}|$, in the Type-I seesaw and minimal seesaw frameworks for the IO scenario with Case I of the CP phases. In contrast to the behaviour of $\Delta m_{21}^{2}$, the RG evolution of $|\Delta m_{32}^{2}|$ is smooth and monotonic throughout the entire energy range considered. As the energy scale evolves from the electroweak scale $\Lambda_{\rm EW}$ to the flavor symmetry scale $\Lambda_{\rm FS}$, the value of $|\Delta m_{32}^{2}|$ decreases continuously in both frameworks, indicating a gradual radiative modification of the atmospheric mass-squared difference.

The comparison between the two seesaw realizations shows that the overall running behaviour is qualitatively similar for all three representative SUSY-breaking scales. For $\Lambda_s=1$ TeV, the Type-I seesaw and minimal seesaw predictions are very close to each other, with the two curves intersecting at an intermediate energy scale. This indicates that the radiative evolution predicted by the two frameworks is nearly identical over most of the energy range. As the SUSY-breaking scale is increased to $\Lambda_s=7$ TeV and $14$ TeV, the minimal seesaw prediction remains slightly larger than the corresponding Type-I seesaw result throughout the RG evolution. However, the numerical difference between the two frameworks remains small, and both models exhibit nearly parallel running behaviour.

These results indicate that the atmospheric mass-squared difference is considerably less sensitive to the differences between the Type-I seesaw and minimal seesaw frameworks than the solar mass-squared difference. Furthermore, the variation of the SUSY-breaking scale has only a minor impact on the relative evolution of the two models. As in the previous analyses, the comparison is performed using the respective phenomenologically viable parameter spaces of the two frameworks. Therefore, the small quantitative differences observed in Fig.~\ref{FIO1} should be interpreted as arising from the combined effects of the distinct model structures and the corresponding input parameter sets, while the overall RG evolution remains qualitatively consistent in both seesaw scenarios.

\textbf{Analyses of R:} For the IO spectrum with Case I of the CP phases, the values of $R_{21}$ exhibit a strong dependence on both the SUSY-breaking scale and the energy scale. At the flavor symmetry scale, $R_{21}$ decreases continuously with increasing $\Lambda_s$, approaching zero for $\Lambda_s=14$ TeV. This behaviour indicates that the predictions of the Type-I seesaw and minimal seesaw frameworks become increasingly similar at higher SUSY-breaking scales. At the electroweak scale, however, the values of $R_{21}$ display a larger variation, reflecting the enhanced sensitivity of the solar mass-squared difference to radiative corrections in the IO spectrum. The corresponding values of $R_{32}$ remain relatively small compared with those of $R_{21}$. Although negative values are obtained at the flavor symmetry scale for all three SUSY-breaking scales, their magnitudes remain below approximately $5\%$, indicating that the atmospheric mass-squared difference exhibits only a modest dependence on the underlying seesaw framework.

Figure~\ref{FR3} presents the relative differences $R_{21}$ and $R_{32}$ between the Type-I seesaw and minimal seesaw frameworks for the IO spectrum with Case I of the CP phases. The blue and orange bars correspond to the values evaluated at the electroweak scale, $\Lambda_{\rm EW}$, and the flavor symmetry scale, $\Lambda_{\rm FS}=10^{14}$ GeV, respectively, for the three representative SUSY-breaking scales, $\Lambda_s=1,\ 7,$ and $14$ TeV.

The left panel illustrates the behaviour of $R_{21}$. At the electroweak scale, the relative difference is largest for $\Lambda_s=1$ TeV, reaching approximately $28.5\%$, indicating a substantial discrepancy between the Type-I seesaw and minimal seesaw predictions for the solar mass-squared difference. As the SUSY-breaking scale increases, the magnitude of $R_{21}$ decreases significantly, changing sign at $\Lambda_s=7$ TeV and becoming positive again with a much smaller value at $\Lambda_s=14$ TeV. This behaviour demonstrates that the relative RG evolution of $\Delta m_{21}^{2}$ is highly sensitive to the SUSY-breaking scale, with the two frameworks exhibiting much closer agreement at larger values of $\Lambda_s$.

At the flavor symmetry scale, $R_{21}$ also decreases steadily with increasing $\Lambda_s$, falling from approximately $6.8\%$ at $\Lambda_s=1$ TeV to a value very close to zero for $\Lambda_s=14$ TeV. The nearly vanishing value of $R_{21}$ at the highest SUSY-breaking scale indicates that the Type-I seesaw and minimal seesaw frameworks yield almost identical predictions for the solar mass-squared difference at the flavor symmetry scale. The right panel presents the corresponding comparison for $R_{32}$. In contrast to $R_{21}$, the relative differences remain comparatively small over the entire parameter space. At the electroweak scale, $R_{32}$ varies between approximately $1.3\%$ and $-3.9\%$, while at the flavor symmetry scale its magnitude remains below about $5\%$ for all three choices of $\Lambda_s$. Although a sign change is observed between $\Lambda_s=1$ TeV and the higher SUSY-breaking scales, the overall magnitude of the relative difference remains modest, indicating that the atmospheric mass-squared difference is only weakly affected by the choice of the seesaw framework.

Overall, Fig.~\ref{FR3} confirms that, for the IO scenario with Case I of the CP phases, the solar mass-squared difference exhibits a considerably stronger dependence on the underlying seesaw realization than the atmospheric mass-squared difference. Moreover, increasing the SUSY-breaking scale generally reduces the discrepancy between the Type-I seesaw and minimal seesaw predictions, particularly for $R_{21}$, where the relative difference approaches zero at the flavor symmetry scale for $\Lambda_s=14$ TeV.

\subsubsection{ Case II of IO}
\textbf{Analyses of $\Delta m_{21}^2$:} The first three panels of Fig.~\ref{FIO2} compare the RG evolution of the solar mass-squared difference, $\Delta m_{21}^{2}$, in the Type-I seesaw and minimal seesaw frameworks for the IO spectrum with Case II of the CP phases. Similar to the Case I scenario, the RG evolution exhibits a non-monotonic behaviour over the intermediate-energy region. However, the evolution patterns in the two seesaw frameworks differ more significantly than those observed in the previous cases.

In the Type-I seesaw framework, $\Delta m_{21}^{2}$ decreases rapidly from the electroweak scale and reaches a pronounced minimum at an intermediate energy scale before increasing monotonically towards the flavor symmetry scale. In contrast, the minimal seesaw framework exhibits a comparatively smoother evolution. Although the value of $\Delta m_{21}^{2}$ also undergoes a moderate decrease after an initial increase, the corresponding minimum is much shallower than that obtained in the Type-I seesaw scenario. Consequently, a sizeable difference between the two frameworks develops over the intermediate-energy region.

The effect of the SUSY-breaking scale is also clearly visible in the comparison. As $\Lambda_s$ increases from $1$ TeV to $14$ TeV, the minimum in the Type-I seesaw evolution becomes deeper, whereas the corresponding minimum in the minimal seesaw framework remains comparatively mild. Nevertheless, both frameworks exhibit a similar increasing trend at higher energies, and the difference between the two predictions gradually decreases as the energy approaches the flavor symmetry scale.

These results indicate that, for the IO spectrum with Case II of the CP phases, the RG evolution of $\Delta m_{21}^{2}$ is considerably more sensitive to the underlying seesaw realization than in the corresponding NO case. The observed differences arise from the combined effects of the distinct theoretical structures of the Type-I seesaw and minimal seesaw frameworks together with the corresponding phenomenologically viable input parameter sets employed in the numerical analysis.

\textbf{Analyses of $\Delta m_{32}^2$:}
The last three panels of Fig.~\ref{FIO2} compare the RG evolution of the atmospheric mass-squared difference, $|\Delta m_{32}^{2}|$, in the Type-I seesaw and minimal seesaw frameworks for the IO spectrum with Case II of the CP phases. In both frameworks, $|\Delta m_{32}^{2}|$ decreases monotonically as the energy scale evolves from the electroweak scale $\Lambda_{\rm EW}$ to the flavor symmetry scale $\Lambda_{\rm FS}$. The smooth evolution of the two curves indicates that the radiative corrections modify the atmospheric mass-squared difference in a gradual and continuous manner throughout the RG running.

The comparison shows that the Type-I seesaw and minimal seesaw frameworks exhibit very similar RG evolution for all three representative SUSY-breaking scales. For $\Lambda_s=1$ TeV, the two curves remain close to each other over the entire energy range, with the minimal seesaw prediction being marginally larger than the corresponding Type-I seesaw result. As the SUSY-breaking scale increases to $\Lambda_s=7$ TeV and $14$ TeV, the overall running behaviour remains essentially unchanged. Although the separation between the two curves becomes slightly more noticeable at lower energies, the difference gradually decreases with increasing energy, and the two predictions become nearly identical in the vicinity of the flavor symmetry scale.

\begin{figure}[t]
    \centering
    \includegraphics[width= \textwidth]{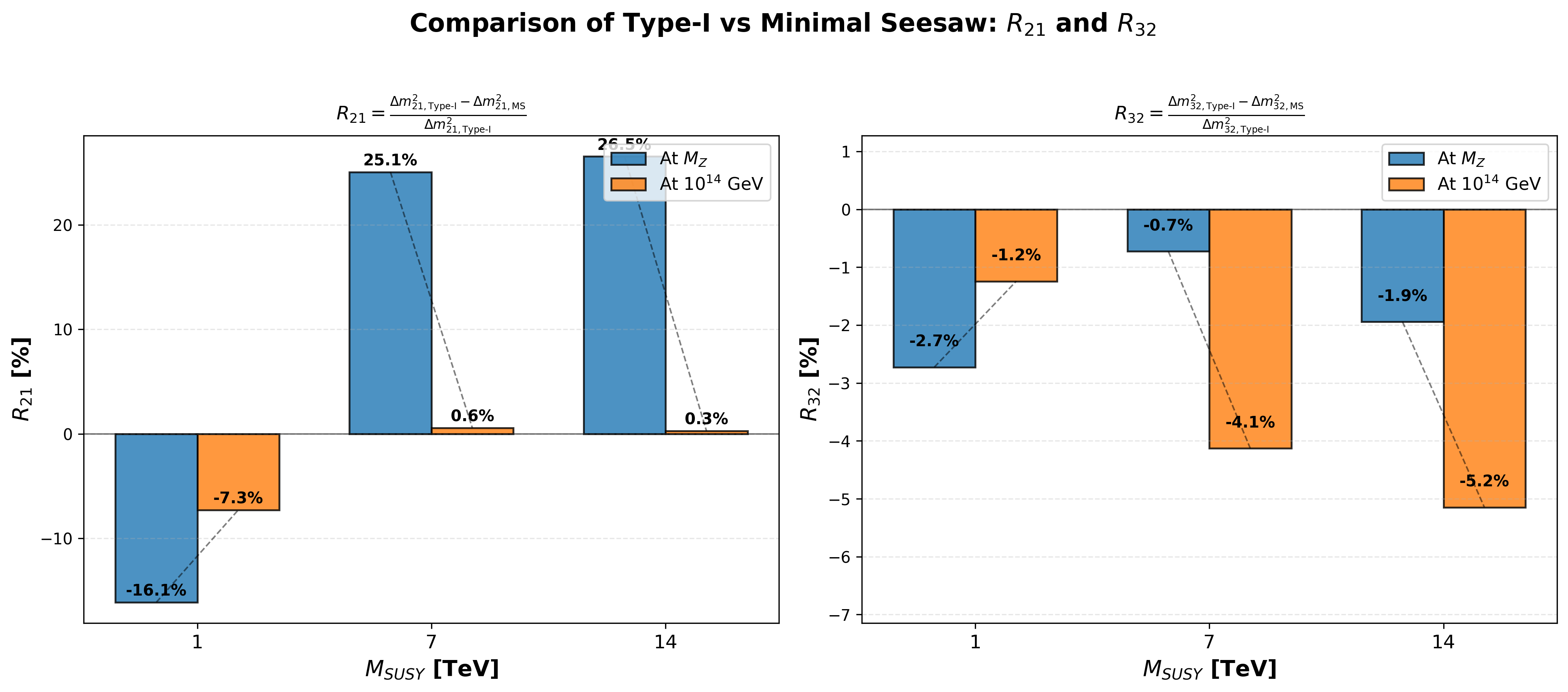}
    \caption{Comparison of $R_{21}$ (left panel) and $R_{32}$ (right panel) between the Type-I seesaw and minimal seesaw frameworks for the IO and Case II. The blue and orange bars correspond to the values at $\Lambda_{\rm EW}$, and the $\Lambda_{\rm FS}=10^{14}$ GeV, respectively, for three SUSY-breaking scales, $\Lambda_s=1,\ 7,$ and $14$ TeV.}
    \label{FR4}
\end{figure}

The results further indicate that the variation of the SUSY-breaking scale has only a minor influence on the relative RG evolution of $|\Delta m_{32}^{2}|$. The qualitative behaviour is preserved for all three choices of $\Lambda_s$, demonstrating that the atmospheric mass-squared difference is comparatively insensitive to both the underlying seesaw realization and the SUSY-breaking scale within the parameter space considered in this work.

As in the previous analyses, the numerical comparison is performed using the respective phenomenologically viable parameter spaces of the Type-I seesaw and minimal seesaw frameworks. Therefore, the small quantitative differences observed in Fig.~\ref{FIO2} arise from the combined effects of the distinct theoretical structures and the corresponding input parameter sets, while the overall RG evolution remains qualitatively similar in the two frameworks.

\textbf{Analyses of R:} For the IO scenario with Case II of the CP phases, the behaviour of $R_{21}$ differs significantly from the previous cases. At $\Lambda_s=1$ TeV, both the flavor symmetry scale and electroweak scale predictions yield large negative values, whereas positive values are obtained for $\Lambda_s=7$ and $14$ TeV. This sign reversal reflects a substantial modification in the relative RG evolution arising from the different CP-phase assignment. Moreover, the magnitude of $R_{21}$ decreases rapidly with increasing $\Lambda_s$, indicating a progressive convergence of the Type-I seesaw and minimal seesaw predictions. The values of $R_{32}$ remain negative for all three SUSY-breaking scales, although their magnitudes are comparatively small. Similar to the other scenarios, the atmospheric mass-squared difference exhibits considerably weaker sensitivity to the underlying seesaw realization than the solar mass-squared difference.

Figure~\ref{FR4} presents the relative differences $R_{21}$ and $R_{32}$ between the Type-I seesaw and minimal seesaw frameworks for the IO spectrum with Case II of the CP phases. The blue and orange bars represent the values of the corresponding ratios evaluated at the electroweak scale, $\Lambda_{\rm EW}$, and the flavor symmetry scale, $\Lambda_{\rm FS}=10^{14}$ GeV, respectively, for the three representative SUSY-breaking scales, $\Lambda_s=1,\ 7,$ and $14$ TeV.

The left panel displays the variation of $R_{21}$. At $\Lambda_s=1$ TeV, both the electroweak-scale and flavor symmetry-scale values are negative, indicating that the minimal seesaw framework predicts larger values of the solar mass-squared difference than the corresponding Type-I seesaw framework. As the SUSY-breaking scale increases to $7$ TeV and $14$ TeV, $R_{21}$ becomes positive at both energy scales. Moreover, while the electroweak-scale values remain close to $25\%$, the corresponding values at the flavor symmetry scale decrease to below $1\%$. This behaviour demonstrates that the two seesaw frameworks become nearly indistinguishable at high energies for larger SUSY-breaking scales, whereas sizeable differences persist in the low-energy predictions. The right panel presents the comparison of $R_{32}$. Unlike $R_{21}$, the relative differences remain negative for all three values of $\Lambda_s$ at both $\Lambda_{\rm EW}$ and $\Lambda_{\rm FS}$. The electroweak-scale values remain relatively small, lying between approximately $-0.7\%$ and $-2.7\%$, whereas the corresponding values at the flavor symmetry scale exhibit a gradual increase in magnitude with increasing $\Lambda_s$, reaching about $-5.2\%$ for $\Lambda_s=14$ TeV. Nevertheless, the overall magnitude of $R_{32}$ remains considerably smaller than that of $R_{21}$, indicating that the atmospheric mass-squared difference is much less sensitive to the underlying seesaw realization.

Overall, Fig.~\ref{FR4} demonstrates that the relative difference associated with the solar mass-squared difference exhibits a much stronger dependence on the SUSY-breaking scale than that associated with the atmospheric mass-squared difference. In particular, the sign reversal of $R_{21}$ between $\Lambda_s=1$ TeV and the higher SUSY-breaking scales highlights the strong influence of the SUSY threshold on the relative RG evolution of the two seesaw frameworks. By contrast, the consistently negative values of $R_{32}$ indicate that the ordering of the atmospheric mass-squared difference predicted by the two frameworks remains unchanged throughout the parameter space considered.

\section{Summary and Conclusion}

The experimentally measured values of the elements of the PMNS mixing matrix provide important clues about the possible existence of an underlying symmetry in the lepton sector. In this context, $\mu$--$\tau$ reflection symmetry has attracted considerable attention, particularly in view of the experimentally preferred nearly maximal value of the atmospheric mixing angle and the indications for a nearly maximal Dirac CP phase. Although such a symmetry can provide a simple explanation for these features, its exact predictions may not remain intact at low energies owing to renormalization group (RG) evolution. The radiative breaking of $\mu$--$\tau$ reflection symmetry therefore provides an interesting framework for studying the deviations of the neutrino parameters from their symmetry predicted maximal values. In this context, the underlying mechanism responsible for neutrino mass generation, such as the conventional Type-I seesaw or its minimal realization, can play an important role in determining the magnitude and pattern of the RG-induced effects.

In the present work, we have investigated the RG-induced breaking of $\mu$--$\tau$ reflection symmetry in the minimal seesaw framework. We have assumed that the symmetry is exact at the high-energy flavor symmetry scale, $\Lambda_{\rm FS}=10^{14}$ GeV, and studied its evolution towards the electroweak scale. The energy dependence of the neutrino parameters then leads to deviations from the symmetry-predicted values at low energies, providing a natural mechanism for radiative breaking of $\mu$--$\tau$ reflection symmetry. Under the exact symmetry, the neutrino parameters are subject to the corresponding symmetry constraints, while their low-energy deviations are determined through the RG evolution.

A central feature of the minimal seesaw framework is that, owing to the rank-two structure of the effective neutrino mass matrix, one of the three light neutrino mass eigenvalues vanishes. Accordingly, we impose $m_1=0$ for the normal ordering and $m_3=0$ for the inverted ordering at the flavor symmetry scale. This reduces the number of independent free neutrino parameters at the high-energy scale from five in the conventional Type-I seesaw framework to four in the minimal seesaw framework. We have therefore investigated whether this reduced set of four free parameters is sufficient to reproduce the experimentally allowed low-energy neutrino data. For both normal and inverted orderings, we have identified suitable sets of high-energy input parameters that successfully reproduce the low-energy neutrino mass-squared differences within the $3\sigma$ ranges of the global-fit data. This demonstrates that the reduced parameter space of the minimal seesaw framework can accommodate the observed low-energy neutrino phenomenology while preserving the assumed high-energy $\mu$--$\tau$ reflection symmetry.

Having established the viability of the minimal seesaw framework, we have further performed a comparative study of its RG evolution with that obtained in the conventional Type-I seesaw scenario. In particular, we have compared the RG running of the solar and atmospheric mass-squared differences, $\Delta m_{21}^{2}$ and $\Delta m_{32}^{2}$, for both normal and inverted mass orderings and for the two considered configurations of the CP phases. To quantify the differences between the two frameworks, we have introduced the relative quantities $R_{21}$ and $R_{32}$ and evaluated them at the electroweak and flavor symmetry scales for different representative values of the SUSY-breaking scale. The corresponding numerical values have been summarized in tables and illustrated through bar diagrams.

The comparison reveals a clear model dependence of the RG evolution. In particular, the relative difference associated with the solar mass-squared difference is generally more pronounced than that associated with the atmospheric mass-squared difference, indicating that $\Delta m_{21}^{2}$ is comparatively more sensitive to the underlying seesaw realization. The magnitude of the difference also depends on the neutrino mass ordering, the CP-phase configuration, and the SUSY-breaking scale. In several scenarios, the predictions of the two frameworks become increasingly close as the SUSY-breaking scale is increased, whereas certain combinations of mass ordering and CP-phase configuration lead to more pronounced deviations. These results demonstrate that the radiative evolution of the neutrino parameters carries information about the underlying mechanism of neutrino mass generation.

Overall, our analysis shows that the minimal seesaw framework, despite its reduced parameter content, can successfully reproduce the observed low-energy neutrino data starting from $\mu$--$\tau$ reflection symmetry at a high-energy scale. The comparison with the conventional Type-I seesaw framework further demonstrates that the RG evolution of neutrino observables is not completely independent of the underlying seesaw realization. The increasing interest in minimal seesaw models, particularly because of their economical particle content and their potential implications for leptogenesis, makes such a comparative study timely. Our results therefore provide a useful perspective on the interplay between $\mu$--$\tau$ reflection symmetry, RG-induced symmetry breaking, and the underlying mechanism of neutrino mass generation.

\end{document}